\documentclass[twocolumn,amsmath,amssymb]{revtex4}
\usepackage{graphicx}
\usepackage{dcolumn}
\usepackage{bm}
\begin{document}

\title{Ultrafast field control of symmetry, reciprocity, and reversibility in buckled graphene-like materials}

\author{Hamed Koochaki Kelardeh\footnote{hkoochakikelardeh1@student.gsu.edu}}
\author{Vadym Apalkov\footnote{vapalkov@gsu.edu}}
\author{Mark I. Stockman\footnote{mstockman@gsu.edu}}
\affiliation{Center for Nano-Optics (CeNO) and
Department of Physics and Astronomy, Georgia State
University, Atlanta, Georgia 30303, USA\\
}

\date{\today}
\begin{abstract}
We theoretically show that buckled two-dimensional graphene-like materials (silicene and germanene) subjected to a femtosecond strong optical pulse can be controlled by the optical field component normal to their plane. In such strong fields, these materials are predicted to exhibit non-reciprocal reflection, optical rectification and generation of electric currents both parallel and normal to the in-plane field direction. Reversibility of the conduction band population is also field- and carrier-envelope phase controllable. There is a net charge transfer along the material plane that is also dependent on the normal field component. Thus a graphene-like buckled material behaves analogously to a field-effect transistor controlled and driven by the electric field of light with subcycle (femtosecond) speed.
\end{abstract}
\maketitle

\section{Introduction}

Novel Dirac materials such as silicene or germanene \cite{review_silicene_2012,graphene_analogous_materials_2013, Effective_lattice_Hamiltonian_MoS2_PRB_2013,
Monolayer_MoS2_PRB_2013,corrugation_Si_Ge_PRB_1994,Quantum_Spin_Hall_effect_PRL_2011,
Low_energy_Hamiltonian_silicene_germanene_PRB_2011, Half_metallic_silicene_germanene_Nano_2012,
Functionalized_silicene_Nano_research_2012} are monolayers of silicon or germanium with 
hexagonal lattice structures where charge carriers at the Fermi surface are, as in graphene,   
Dirac fermions \cite{silicene_APL_2010,silicene_APL_2011, silicene_exp_PRL_2012_Vogt, Evidence_silicene_PRL_2012,
Structure_silicene_applied_physics_express_2012, evidence_silicene_Nano_letters_2012, Silicene_on_diboride_PRL_2012, germanene_experimental_New_J_Physics_2014, Germanene_on_Pt_advanced_materials_2014}. Recently, silicene has shown  \cite{Tao_et_al_nnano.2014.325_2015_Silicine_FET} promise for applications in electronics such as field-effect transistors (FETs) \cite{Kahng_MOSFET_Patent_1963, Taur_Ning_1998_VLSI_Fundamentals, Liou_Schwierz_2003_Microwave_Transistors, Schwierz_Wong_Liou_2010_Nanometer_CMOS} where, being a semiconductor, it has a natural advantage over graphene that is a semimetal. Below we will consider silicene but all qualitative results are also valid for germanene.

\begin{figure}
\begin{center}\includegraphics[width=0.48\textwidth]{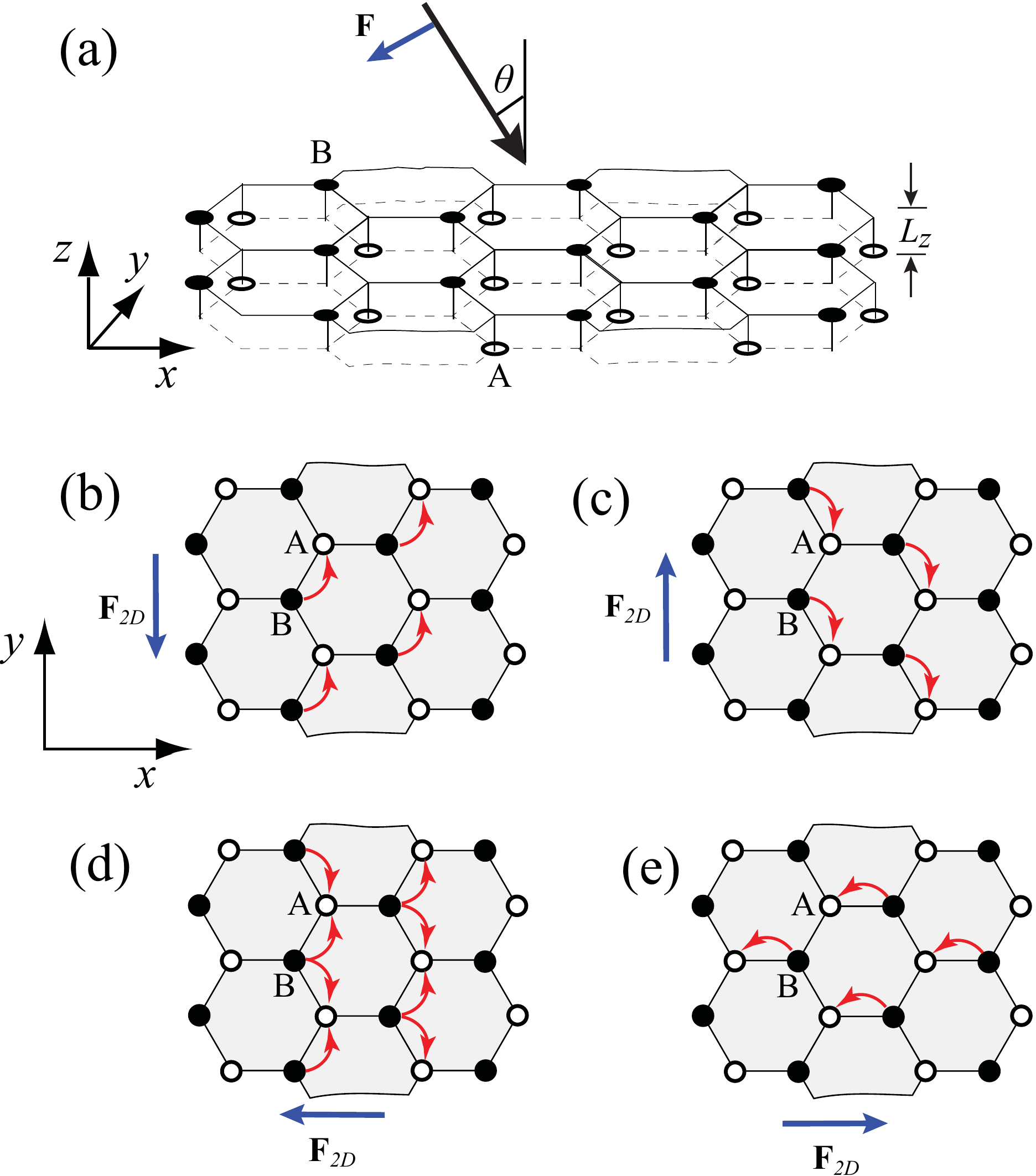}\end{center}
\caption{(a) Hexagonal lattice structure of 2D silicene. The lattice consists of two 
inequivalent sublattices labeled by ``A" and ``B". 
Sublattices A and B are shifted in the $z$ direction by distance $L_z$. The angle of incidence of the 
pulse is $\theta $. 
(b) Schematic of in-plane electron transfer induced by in-plane pulse electric field $\mathbf{F}_{2D}$ directed along the $y$ axis as shown. The curved red arrows indicate the electron transfer between the sublattices. (c) The same as (b) but for the opposite $\mathbf{F}_{2D}$. (d) The same as (b) but for the field directed along the $x$ axis. (e) The same as (d) but for the opposite $\mathbf{F}_{2D}$. The $z$ components of the pulse field has the same direction in all cases.
} 
\label{silicene5.pdf}
\end{figure}

In this paper we theoretically predict that a single monolayer of silicene (germanene) is controllable at optical frequencies by a normal component of the incident optical field just like the gate voltage controls channel current in FET. The main difference between silicene and graphene is that due to a larger radius of a Si (or, Ge in germanene) atom compared to a C atom, the 
corresponding hexagon lattice in silicene has {\it 
buckled} structure \cite{buckled_silicene_on_Ir_Nano_2012} consisting of two sublattices that are displaced vertically by a finite distance $L_z \sim 0.5~\mathrm{\AA}$ -- see Fig.\ \ref{silicene5.pdf}(a).
As a result,  silicene has large spin-orbit interaction, which opens up band gaps at the Dirac points ($\Delta^{}_{\rm so} \approx 1.55 - 7.9$ meV for silicene 
\cite{Quantum_Spin_Hall_effect_PRL_2011, Silicene_effective_Hamiltonian_PRB_2011} and $\Delta^{}_{\rm so} \approx 24 - 93$ meV for germanene \cite{Quantum_Spin_Hall_effect_PRL_2011, Silicene_effective_Hamiltonian_PRB_2011}). For graphene, 
the corresponding spin-orbit-induced gap is very small, 25 $\mu$eV \cite{Band_structure_topologies_graphene_PRB_2009}. 
The buckled structure of silicene/germanene lattice allows also for the band gap to be controlled by an applied perpendicular 
electric field \cite{silicene_inhomogeneous_field_New_J_phys_2012}:  the band gap increases almost linearly with this electric field. 

Phenomena in silicene in a strong optical pulse field are illustrated in  Figs.\ \ref{silicene5.pdf}(b)-(e). A strong optical field causes electron transfer in the direction of the force \cite{Schiffrin_at_al_Nature_2012_Current_in_Dielectric, Stockman_et_al_PRB_2015_Graphene_in_Strong_Field}. In fact, a strong optical field in the $z$-direction (normal to the silicene plane) decreases symmetry of the system from honeycomb (six-order, centrosymmetric)  to triangular (third-order, non-centrosymmetric). This leads to appearance of effects such as optical rectification and induction of currents normal to the in-plane component of the applied electric field.
 
Microscopically, the $z$ component of the strong field causes transfer of electrons between the sublattices. Assume for certainty that, for the chosen pulse, electrons are transferred from A to B. (Note that the change of the maximum field to the opposite, i.e., change of the carrier-envelope phase of the pulse by $\pi$, would obviously cause an opposite transfer.) In the case of in-plane field $\mathbf{F}_{2D}$ polarized in the $y$-direction, there is an electron transfer in both the $y$- and $x$-directions -- see Fig.\ \ref{silicene5.pdf}(b). The symmetry of the system dictates that with the reversal of $\mathbf{F}_{2D}$ (for the same $z$-component, $F_z$) the
$y$-current changes to the opposite but the $x$-current does not change, as shown in Fig.\ \ref{silicene5.pdf}(c). This implies, in particular, that the system causes optical rectification in the $x$-direction, which is due to the absence of symmetry with respect to the reflection in $yz$-plane for either sublattices.

Fundamentally different scenario takes place for $\mathbf{F}_{2D}$ in the $x$ direction -- see Figs.\ \ref{silicene5.pdf}(d) and (e).  In this case, there is no current in the $y$-direction due to symmetry with respect to reflection in the $xz$-plane. With respect to field $\mathbf{F}_{2D}$ changing to the opposite, the $x$-current does not have any definite parity, which is rectification in the $x$-direction.

To provide for the field-effect control of optical phenomena in silicene, the $z$-component of the  pulse electric field should be strong enough: $F_z\gtrsim \hbar\omega/(e L_z)\sim 2~\mathrm{V/\AA}$, where $\omega$ is the optical frequency. Then, necessarily, the pulse should be very short, on the femtosecond scale, to allow the processes to be complete before significant damage to the lattice may have occurred -- see Sec.\ \ref{Model} below. For such fields, there may be partial adiabaticity (reversibility) set on, which we will show below in Sec.\ \ref{Results}.

\section{Model and Main Equations}
\label{Model}

At the present time, record-setting ultrashort optical pulses have duration $\approx 1.5$ optical period \cite{Schiffrin_at_al_Nature_2012_Current_in_Dielectric, Schultze_et_al_Nature_2012_Controlling_Dielectrics}, with 
 duration of just a few femtoseconds. We will idealize and simulate such an ultrashort pulse with the following single-oscillation waveform,  
\begin{equation}
F (t) = F_{0} e^{-u^2} \left( 1 - 2 u^2 \right),
\label{FV0}
\end{equation}
where $F_0$ is the amplitude, which is related to the pulse power, ${\cal P} =c F_0^2/4 \pi$,  
$c$ is speed of light,
$u = t/\tau $, and $\tau $ is the pulse length, which is set $\tau = 1 $ fs. Note that this waveform has zero area, $\int_{-\infty}^\infty F(t)dt=0$, which is required for a pulse propagating in far-field zone.

We consider a $p$-polarized laser pulse with polarization direction parallel to the plane of 
incidence, orientation of which is determined by an angle $\varphi $ measured relative to axis $x$. 
 Here the $xy$ coordinate system is introduced in the plane of silicene/germanene, oriented as shown in Fig.\ \ref{silicene5.pdf}(a). The angle of incidence of the laser pulse is denoted as $\theta $. 
 
\begin{figure}
\begin{center}\includegraphics[width=0.48\textwidth]{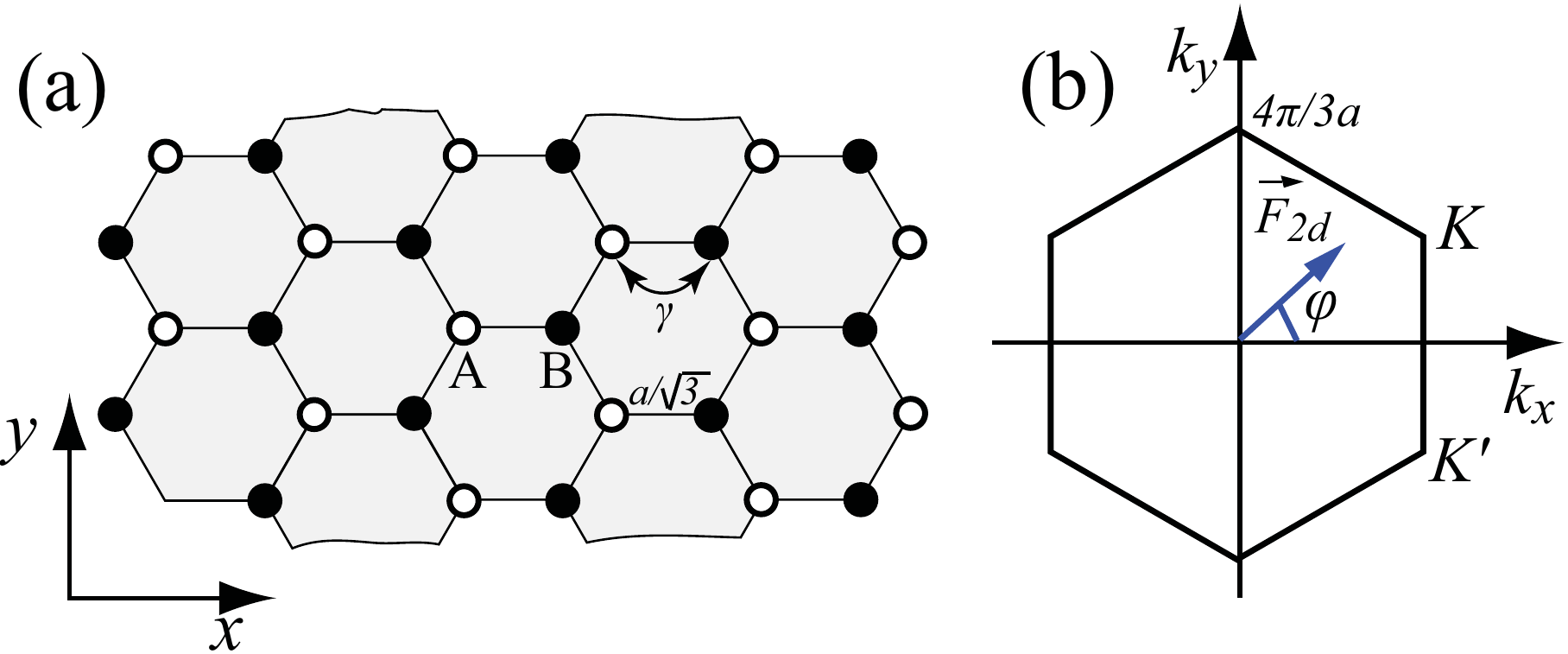}\end{center}
\caption{(a) Hexagonal lattice structure of 2D silicene/germanene. The graphene lattice consists of two 
inequivalent sublattices labeled by A and B. The vectors $\mathbf{a}_1=a/2(\sqrt{3},1)$ and $\mathbf{a}_2 = a/2(\sqrt{3},-1) $ are the direct lattice vectors of silicene/germanene. The nearest neighbor coupling, which is characterized by the hopping integral $\gamma $, is also shown. 
(b) The first Brillouin zone of silicene/germanene. Points $K$ and $K^{\prime }$ are two degenerate Dirac points, corresponding to two valleys of low energy spectrum of silicene/germanene. 
The blue arrow shows in-plane ($xy$-plane) component of the time-dependent electric field of the pulse. The in-plane field, $\mathbf{F}_{2d}$, is characterized by azimuthal angle $\varphi $.
} 
\label{silicene2.pdf}
\end{figure}

Similar to graphene, the silicene/germanene monolayer has honeycomb lattice structure, which is 
shown in Fig.\ \ref{silicene2.pdf}(a). The lattice has two sublattices, labeled ``A" and ``B", and is determined by two lattice vectors,  $\mathbf{a}_1=a/2(\sqrt{3},1)$ and $\mathbf{a}_2 = a/2(\sqrt{3},-1) $, where 
$a$  is lattice constant, which is 
$3.866$ \AA \ for silicene and $4.063$ \AA \ for germanene. The distance between the nearest neighbors is $a/\sqrt{3}$. The first Brillouin zone of the reciprocal lattice is a hexagon and is shown in Fig.\ \ref{silicene2.pdf}(b). The points $K= (2\pi/a) (1/\sqrt{3},1/3)$ and $K^{\prime}= (2\pi/a) (1/\sqrt{3},-1/3)$ are the Dirac points. 
In the buckled structure [see Fig.\ \ref{silicene5.pdf}(a)], the $z$-shift distance is $L_z=0.46$ \AA \  and $L_z=0.66$ \AA \ for silicene and germanene, respectively \cite{Liu_Jiang_Yao_PhysRevB.84_2011_Spin_Orbit_in_Germanium_and_Tin, Liu_Feng_Yao_PhysRevLett.107_2011_Quantum_Spin_Hall_Effect_in_Solicene}.

For a graphene monolayer, where spin-orbit coupling is extremely small ( $\approx 0.03$ meV),  energy gaps at the Dirac points are correspondingly very small and can be set as zero for any practical purposes. Then the low energy spectra near the Dirac points are well described by the Dirac massless relativistic equation. For a silicene/germanene system, finite spin-orbit interaction opens up a much larger gap $\sim10-100$ meV \cite{silicene_inhomogeneous_field_New_J_phys_2012}. Such a gap in the energy spectrum of silicene modifies  low-energy electron transport and  interaction between electrons in weak magnetic fields \cite{Tunability_FQHE_dirac_materials_PRB_2014}. However, this spin-orbit interaction is too weak and, in our case,  can be safely neglected compared to characteristic energy scale, $eF_zL_z$, introduced by the  strong electric field of the optical pulse in the buckled Dirac materials.  At the same time, the buckled structure of a silicene monolayer introduces strong sensitivity of the system to the external normal field, $F_z$ \cite{silicene_inhomogeneous_field_New_J_phys_2012}.  Hence, based on this consideration, below in this article we disregard  the  spin-orbit interaction but take into account the buckled structure bringing about the sensitivity to the normal optical electric field.

The Hamiltonian of an electron in silicene in the field of an optical pulse has the form 
\begin{equation}
{\cal H} = {\cal H}_0 + e \mathbf{F}_{2d}(t)\mathbf{r} + \frac{e L_zF_z(t)}{2}\left(
\begin{array}{cc}
 1 &  0 \\
0  & -1
\end{array}
\right)   ,
\label{Htotal}
\end{equation}
where ${\cal H}_0$ is the field-free electron Hamiltonian, $\mathbf{r} = (x,y)$ is a two dimensional vector, 
$\mathbf{F}_{2d} = (F_x(t), F_y(t)) = F(t)\sin \theta (  \cos\phi  , \sin\phi )$,  and 
$F_z(t) = F(t) \cos \theta$. 
Here the matrix form of the Hamiltonian corresponds to pseudo-spin, i.e., two components of the wave function  $\psi_A$ and $\psi_B$,
which describe the amplitudes for an electron to be on the lattice site $A$ and $B$, respectively. 

The field-free electron Hamiltonian, ${\cal H}_0$, describes 
 the nearest neighbor tight-binding model of silicene without spin-orbit terms. This 
 Hamiltonian is exactly the same as the free-field Hamiltonian of graphene \cite{Graphene_Wallace_PR_1947,Graphene_Weiss_PR_1958, graphene_Dresselhaus_1998,Carbon_nanotubes_2004} and 
describes the tight-binding 
coupling between two sublattices A and B -- see Fig.\ \ref{silicene2.pdf}(a). In the reciprocal space, the Hamiltonian 
${\cal H}_0$ is a $2\times 2$ matrix of the form \cite{Graphene_Wallace_PR_1947,Graphene_Weiss_PR_1958}
\begin{equation}
{\cal H}_0 =  
\left(
\begin{array}{cc}
0  &  \gamma  f(\mathbf{k}) \\
\gamma f^{*}(\mathbf{k})  & 0
\end{array}
\right)   ,
\end{equation}
where the hopping integral $\gamma$  is $-1.6$ eV for silicene and $-1.3$ for germanene \cite{silicene_inhomogeneous_field_New_J_phys_2012}, and 
\begin{equation}
f(\mathbf{k}) = \exp \left( i \frac{a k_x} {\sqrt{3}}\right) + 2  \exp \left( -i \frac{a k_x} {2 \sqrt{3}}\right) 
 \cos \left( \frac{ak_y }{2} \right).
\end{equation}

The energy spectrum of Hamiltonian ${\cal H}_0$ consists of the conduction band (CB) ($\pi^*$, or anti-bonding band) and the valence band (VB) ($\pi$, or bonding band) with energy 
dispersion $E_{c}(\mathbf{k})= -\gamma |f(\mathbf{k})|$ (CB) and $E_{v}(\mathbf{k})= \gamma |f(\mathbf{k})|$ (VB). The corresponding 
wave functions are 
\begin{equation}
\Psi^{(c)}_{\mathbf{k}} (\mathbf{r} ) = \frac{e^{i \mathbf{k} \mathbf{r}}}{\sqrt{2}}
\left( 
\begin{array}{c}
1 \\
e^{-i \phi_k }
\end{array}
\right)  
\label{functionV}
\end{equation}
and
\begin{equation}
\Psi^{(v)}_{\mathbf{k}} (\mathbf{r} ) = \frac{e^{i \mathbf{k} \mathbf{r}}}{\sqrt{2}}
\left( 
\begin{array}{c}
-1 \\
e^{-i \phi_k }
\end{array}
\right)  ,
\label{functionC}
\end{equation}
where $ f(\mathbf{k}) = |f(\mathbf{k}) | e^{i\phi _k}$. 

The characteristic electron-electron scattering time $\tau_{e-e}$ in silicene/germanene is expected to be similar to 
the corresponding time is graphene, which is  $\sim10-100$ fs \cite{Hwang_Das_Sarma_PRB_2008_Graphene_Relaxation_Time, ultrafast_dynamics_graphene_PRB_2011,
theory_absorption_ultrafast_kinetics_graphene_PRB_2011,
Ultrafast_collinear_scattering_graphene_nat_comm_2013,non-equilibrium_Dirac_carrier_distributions_graphene_nat_materials_2013,
Nonequilibrium_dynamics_photoexcited_electrons_graphene_PRB_2013}. The duration of the pulse in 
our problem ($\tau_p\sim 4$ fs) is  $\tau_p\lesssim\tau_{e-e}$. Therefore, it is not unreasonable to assume that 
the electron dynamics in the external 
electric field of the optical pulse is coherent and can be described 
by time-dependent Schr\"odinger equation
\begin{equation}
i\hbar \frac{d \Psi}{dt} = {\cal H} \Psi,
\label{time_eq1}
\end{equation} 
where Hamiltonian ${\cal H}$ of Eq.\ (\ref{Htotal}) has an explicit time dependence. 

The electric field of the optical pulse generates both interband and intraband electron dynamics. The 
interband dynamics introduces coupling of the states of the CB and VB and results in 
redistribution of electrons between the two bands. For dielectrics, such dynamics results in its metallization, which manifests itself as a finite charge transfer through dielectrics and finite CB population after the 
pulse ends \cite{Schiffrin_at_al_Nature_2012_Current_in_Dielectric, Krausz_Stockman_Nature_Photonics_2014_Attosecond_Review, Schultze-2014-Attosecond_band-gap}.
 
In the reciprocal space, the intraband dynamics is described by the acceleration theorem \cite{Wannier_PR_1960_Wannier_States_in_Strong_Fields},
\begin{equation}
\hbar \frac{d\mathbf{k} }{dt} = e \mathbf{F}(t).  
\label{acceleration}
\end{equation}
This acceleration theorem is universal and does not depend on the dispersion law. Therefore the intraband electron dynamics is 
the same for both the VB and CB. The time-dependent wave vector $\mathbf{k}_T(\mathbf{q},t)$ of an electron with initial wave vector  $\mathbf{q}$ can be found by solving Eq.\ (\ref{acceleration}) as
\begin{equation} 
\mathbf{k}_T(\mathbf{q},t) = \mathbf{q} +  
\frac{e}{\hbar} \int^t_{-\infty}  \mathbf{F} (t_1) dt_1 .
\label{kT}
\end{equation}
The corresponding electron wave functions are the well-known Houston functions
\cite{Houston_PR_1940_Electron_Acceleration_in_Lattice},
\begin{equation}
\Phi^{(H)}_{\alpha \mathbf{q} }(\mathbf{r},t) = \Psi^{(\alpha)}_{\mathbf{k}_T(\mathbf{q},t)} (\mathbf{r} )
e^{- \frac{i}{\hbar }   \int ^t_{-\infty} \!\! dt_1 E_{ \alpha} [\mathbf{k}_T(\mathbf{q},t_1) ] }, 
\label{phi_h}
\end{equation}
where $\alpha = v$ (VB) or $\alpha = c$ (CB). 

Using the Houston functions as a basis, we express the general solution of time-dependent
Schr\"odinger equation (\ref{time_eq1}) in the following form
\begin{equation}
\Psi_{\mathbf{q}} (\mathbf{r}, t) = \sum_{\alpha = v,c} \beta _{\alpha \mathbf{q} } (t) 
                  \Phi^{(H)}_{\alpha \mathbf{q} }(\mathbf{r},t).
 \label{psi0}
\end{equation}
Solution (\ref{psi0}) is parametrized by initial electron wave vector $\mathbf{q}$. Due to the
universal intraband electron dynamics in the reciprocal space, the equations, which describe 
coherent electron dynamics in the pulse field, become decoupled,  greatly 
simplifying the problem. 

Expansion coefficients $\beta _{\alpha \mathbf{q} }$ satisfy the following system of differential equations 
\begin{eqnarray}
&&\frac{d \beta _{c \mathbf{q} } (t)}{dt} =
\nonumber\\
&& -i \frac{\mathbf{F}_{2d}(t) \mathbf{Q} _{\mathbf{q}} (t) + e F_z(t) 
                                                   \tilde{L}_z(t, \mathbf{q})}{\hbar } 
  \beta _{v \mathbf{q} } (t),    \label{system1} \\
& & 
\frac{d \beta _{v \mathbf{q} } (t)}{dt} = 
\nonumber\\
&&-i \frac{\mathbf{F}_{2d}(t)  \mathbf{Q}^* _{\mathbf{q}} (t) + e F_z(t) 
                                                   \tilde{L}_z (t, \mathbf{q}) }{\hbar } 
  \beta _{c \mathbf{q} } (t),
\label{system2}
\end{eqnarray}
where function $\tilde{L}_z (t, \mathbf{q}) $, which is given by the following expression,
\begin{equation}
\tilde{L}_z (t, \mathbf{q}) 
 =  L_z  
   e^{- \frac{i}{\hbar }   \int ^t_{-\infty} \!\! dt_1 
       \left\{  E_{c} [\mathbf{k}_T(\mathbf{q},t_1)]  -E_{v} [\mathbf{k}_T(\mathbf{q},t_1)]  \right\}  },
\label{Lztime}
\end{equation}
is specific to the buckled structure of silicene. It determines the interband coupling induced by the
perpendicular component of the pulse electric field. Vector function $\mathbf{Q} _{\mathbf{q}} (t)$ is proportional to the in-plane interband dipole matrix element,
\begin{equation}
\mathbf{Q} _{\mathbf{q}} (t) =  \mathbf{D}[\mathbf{k}_T(\mathbf{q},t)]   
   e^{- \frac{i}{\hbar }   \int ^t_{-\infty} \!\! dt_1 
       \left\{  E_{c} [\mathbf{k}_T(\mathbf{q},t_1)]  -E_{v} [\mathbf{k}_T(\mathbf{q},t_1)]  \right\}  },
\label{Q_vector}
\end{equation}
where $\mathbf{D}(\mathbf{k} ) =\left(D_x(\mathbf{k}),D_y(\mathbf{k})\right)  $ is the dipole matrix element between the states of the CB and  
VB with the same wave vector $\mathbf{k}$, namely,
\begin{equation}
\mathbf{D}(\mathbf{k} ) = \left\langle \Psi^{(c)}_{\mathbf{k} }\right|  e\mathbf{r} 
\left| \Psi^{(v)}_{\mathbf{k} }\right\rangle.
\label{dipole}
\end{equation}
Substituting Eqs.\ (\ref{functionV}) and 
(\ref{functionC}) into Eq.\ (\ref{dipole}), we obtain explicitly,
\begin{equation}
D_x(\mathbf{k})  = \frac{e a}{2\sqrt{3}}  \frac{1+ \cos \left( \frac{ak_y}{2}  \right) 
       \left[ \cos \left(\frac{3ak_x}{2\sqrt{3} } \right) - 2\cos \left(\frac{ak_y}{2}  \right)   \right]  }
         {1+4 \cos \left( \frac{ak_y}{2}  \right) 
       \left[ \cos \left(\frac{3ak_x}{2\sqrt{3} } \right) + \cos \left(\frac{ak_y}{2}  \right)   \right]  } 
\label{Dx}
\end{equation}
and 
\begin{equation}
D_y(\mathbf{k})  = \frac{e a}{2}  \frac{\sin \left( \frac{ak_y}{2}  \right)  
      \sin \left(\frac{3ak_x}{2\sqrt{3} } \right) }
         {1+4 \cos \left( \frac{ak_y}{2}  \right) 
       \left[ \cos \left(\frac{3ak_x}{2\sqrt{3} } \right) + \cos \left(\frac{ak_y}{2}  \right)   \right]  }. 
\label{Dy}
\end{equation}
System of equations (\ref{system1})-(\ref{system2}) describes the interband electron dynamics and determines the mixing of CB and VB states in the electric field of the pulse.  
For undoped silicene, all VB states  are initially occupied 
and all CB states are empty. Then the initial condition for system Eqs.\ (\ref{system1})-(\ref{system2}) is $(\beta _{v \mathbf{q} }, \beta _{c \mathbf{q} } )  = (1,0)$, and the mixing of the states 
of different bands is characterized by time-dependent component  $|\beta _{c \mathbf{q} }(t)|^2$. We 
also define the time-dependent total CB population  by the 
following expression, 
\begin{equation}
{\cal N}_c (t) = \sum_{\mathbf{q}} |\beta _{c \mathbf{q} }(t)|^2,
\label{Ntotal}
\end{equation}
where the sum is over the first Brillouin zone. The CB population, ${\cal N}_c (t)$, characterizes the 
electron dynamics in silicene and determines whether  the dynamics for the entire  system is reversible 
or not. Namely, the dynamics is reversible if, after the pulse ends, the CB population, 
which is the residual CB population, is small compared to the maximum CB population 
throughout the pulse. 

Polarization of the system in a time-dependent electric field also generates electric current, 
which can be calculated in terms of the velocity operator
from the following expression 
\begin{equation}
J_j (t) = \frac{e}{a^2} \sum_{\mathbf{q}} \sum_{\alpha_1 =v,c} 
\sum_{\alpha _2 = v,c}
\beta _{\alpha_1 \mathbf{q}}^* (t) 
    {\cal V}_{j}^{\alpha_1 \alpha_2}  \beta _{\alpha_2\mathbf{q}} (t),
\label{current}
\end{equation}
where $j = x,y$, and $ {\cal V}_{j}^{\alpha_1 \alpha_2} $ are 
matrix elements of the velocity operator 
$\hat {\cal V}_{j} = \frac{1}{\hbar } 
\frac{\partial {\cal H}_0}{ \partial k_j}$. With the known wave functions (\ref{functionV})-(\ref{functionC}) of the 
CB and VB, the matrix elements of the velocity 
operator are 
\begin{eqnarray}
 {\cal V}_x^{cc} = -{\cal V}_x^{vv} =& & \frac{a\gamma }{\sqrt{3} \hbar} 
\left[  \sin\left( \frac{ak_x}{\sqrt{3}} - \phi_{\mathbf{k}}   \right)
\right.  +  \nonumber \\
& & \left.   \sin\left( \frac{ak_x}{\sqrt{3}} + \phi_{\mathbf{k}}   \right)  \cos \frac{ak_y}{2} \right],
\end{eqnarray}
\begin{equation}
{\cal V}_y^{cc} = -{\cal V}_y^{vv} = \frac{a\gamma}{\hbar}    \cos\left( \frac{ak_x}{2\sqrt{3}} + \phi_{\mathbf{k}}   \right)   \sin \frac{ak_y}{2} ,
\end{equation}
\begin{eqnarray}
{\cal V}_x^{cv} =  -i \frac{2a\gamma}{\sqrt{3} \hbar} & & 
\left[   \cos\left( \frac{ak_x}{\sqrt{3}} - \phi_{\mathbf{k}}   \right)  -
\right.   \nonumber \\
& &  \left. \cos\left( \frac{ak_x}{\sqrt{3}} + \phi_{\mathbf{k}}   \right)  \cos \frac{ak_y}{2} \right],
\end{eqnarray}
and
\begin{equation}
{\cal V}_y^{cv} =  -i \frac{2a\gamma}{\hbar}  \sin\left( \frac{ak_x}{\sqrt{3}} + \phi_{\mathbf{k}}   \right)  \cos \frac{ak_y}{2}. 
\end{equation}
The interband matrix elements of the velocity operator, ${\cal V}_x^{cv}$ and 
${\cal V}_y^{cv}$, are related to the interband dipole matrix elements, 
${\cal V}_x^{cv} = i D_x(\mathbf{k})  \left[  E_c (\mathbf{k}) - E_v (\mathbf{k})  \right]/\hbar $ and ${\cal V}_y^{cv} = i D_y(\mathbf{k})  \left[  E_c (\mathbf{k}) - E_v (\mathbf{k})  \right]/\hbar $ \cite{Landau_Lifshitz_Quantum_Mechanics:1965}.
Within the nearest-neighbor tight binding model, silicene has electron-hole symmetry, which results in the relation ${\cal V}_y^{cc} = -{\cal V}_y^{vv}$.

Let us denote current in the $i$ direction induced by in-plane field $\mathbf{F}_{2d}$ in the $j$ direction as $J_{ij}$, where $i,j=x,y$. Similarly we denote charge transferred after the pulse ends through the system as $Q_{ij}$. This is determined by an expression 
\begin{equation}
Q_{ij} = \int_{-\infty}^{\infty }  dt J_{ij} (t),
\label{Qtr}
\end{equation}
The current can be expressed in terms of polarization $\mathbf{P}(t)$ of the electron system as $\mathbf{J}(t)= d\mathbf{P}(t)/dt$. Then the transferred charge is determined by the residual polarization of the system as $Q_{ij} = P_{ij}(t\rightarrow \infty )$, where we introduced tensor indices for $P$ similarly to those for $J_{ij}$ and $Q_{ij}$. The transferred charge is nonzero only due to irreversibility of electron dynamics in the optical pulse field. For completely reversible dynamics, when the system returns to its initial state after the pulse, the transferred charge would be exactly zero. 

\section{Results and Discussion}
\label{Results}

\subsection{Band Population Dynamics in Strong Pulse Field}
\label{Populations}

\begin{figure}
\begin{center}\includegraphics[width=0.48\textwidth]{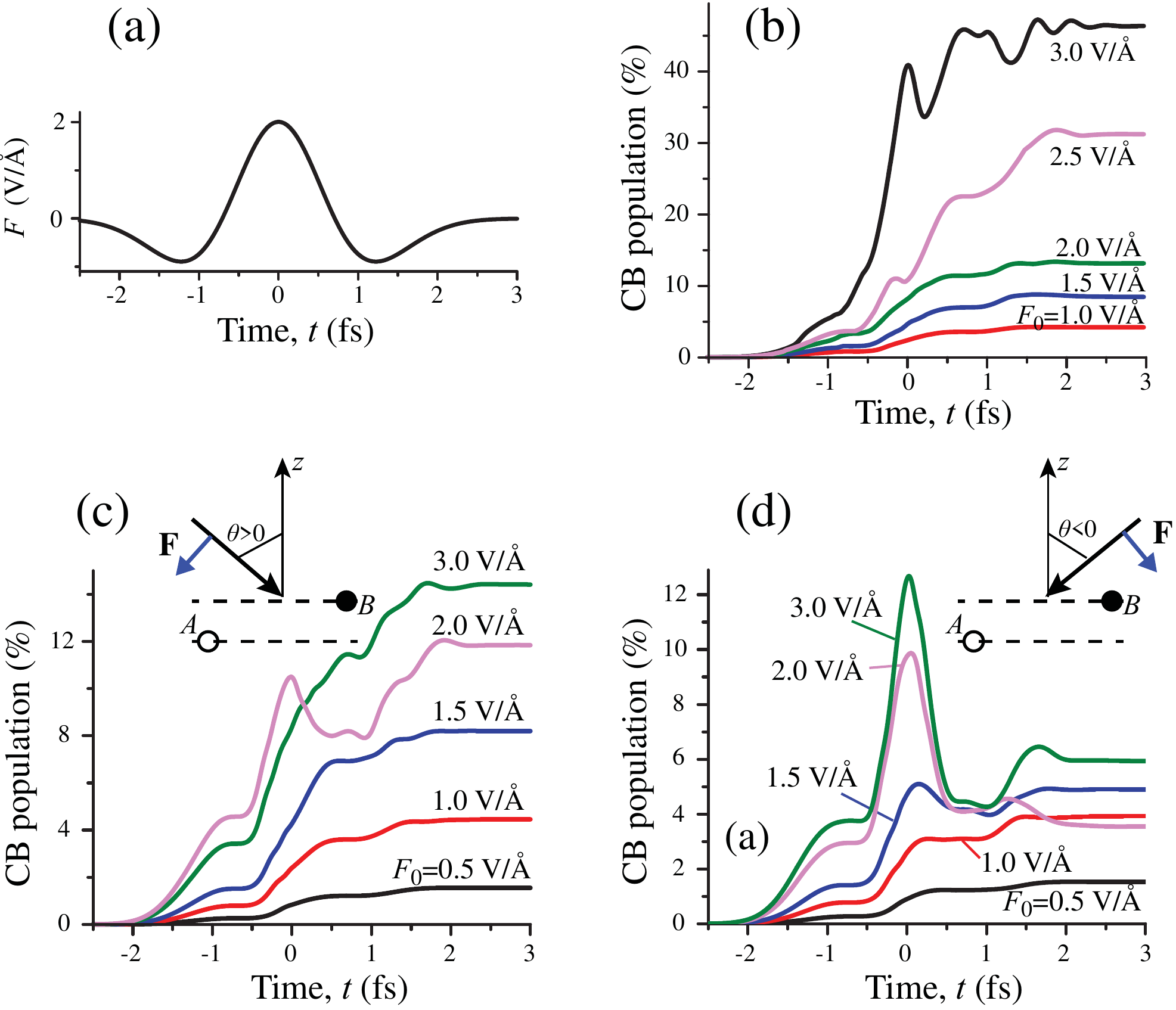}\end{center}
\caption{(a) Pulse waveform as given by Eq.\ (\ref{FV0}) for $F_0=2~\mathrm{V/\AA}$. (b) For excitation pulse polarized in the $yz$ plane, CB population ${\cal N}_c$ is shown as a function of time for pulse amplitudes $F_0$ indicated. Incidence angle $\theta=\pm80^\circ$.
(c) The same as (b) but for the pulse polarized in the $xz$ plane with the direction of the maximum field, $\mathbf F$, shown in the inset; incidence angle $\theta=80^\circ$. (d) The same as (c) but for $\theta=-80^\circ$.
} 
\label{populations_silicene_80_deg.pdf}
\end{figure}

The principal distinction of silicene from graphene is that the sublattices, A and B, are separated ``vertically" (i.e., in the $z$-direction) by an appreciable distance, $L_z\approx 0.5~\mathrm\AA$ -- see Fig.\ \ref{silicene5.pdf}(a). The strong field of the optical pulse causes non-perturbative nonlinear changes in the material. Such phenomena are sensitive to the maximum field of the pulse, which is amplitude $F_0$. For our choice of pulse Eq.\ (\ref{FV0}), the maximum of the carrier oscillation occurs at the maximum of the pulse envelope, i.e., the carrier-envelope phase (CEP) is zero -- see Fig.\ \ref{populations_silicene_80_deg.pdf}(a) illustrating the pulse waveform. 

The CB population, ${\cal N}_c$, calculated in accord with Eq.\ (\ref{Ntotal}) for pulse polarized in the $yz$ plane  is displayed in Fig.\ \ref{populations_silicene_80_deg.pdf}(b) as a function of time $t$ for different field amplitudes and incidence angle $\theta=\pm80^\circ$. Note that because silicene is symmetric with respect to reflection in the $xz$ plane, the results for both $80^\circ$ and $-80^\circ$ are identical. Two most prominent features of this dynamics are: {\it(i)} dependence on the pulse amplitude is very nonlinear,  and {\it(ii)} the residual (after the pulse end) populations,  ${\cal N}_c^{\mathrm(res)}$ are close to the maximum populations during the pulse. The latter property is similar to that in graphene \cite{Stockman_et_al_PRB_2015_Graphene_in_Strong_Field}.
However, it is in a sharp contrast to that in silica, cf.\ Refs.\ \cite{Apalkov_Stockman_PRB_2012_Strong_Field_Reflection} (theory) and \cite{Schiffrin_at_al_Nature_2012_Current_in_Dielectric, Schiffrin_at_al_Nature_2014_Current_in_Dielectric_Addendum} (experiment) where the residual CB populations are relatively small. This large residual CB population for silicene suggests lack of adiabaticity, which is likely due to a relatively small distance of the transfer between the two sublattices in the $xy$ plane, $L_{xy}=a/\left(2\sqrt{3}\right)\approx 0.7~\mathrm\AA$, in this case. Note that the adiabaticity parameter is $\delta= \hbar\omega/\left(eF_y L_{xy}\right)$. Adiabaticity requires $\delta\ll1$ while, in our case, even at the strongest fields, the adiabatic parameter is not too small, $\delta \gtrsim 1$.

The response for the case of the pulse polarized in the $xz$ plane is displayed in Figs.\ \ref{populations_silicene_80_deg.pdf}(c)-(d). In the stark contrast to the case of the $yz$ polarization considered above in the previous paragraph, here there is a dramatic difference between $\theta=80^\circ$ and $\theta=-80^\circ$. This is due to the violation in the reflection symmetry induced by the $z$ component of the maximum field. For the case illustrated, this field promotes transfer of electrons predominantly toward the B sublattice -- cf.\ Fig.\ \ref{silicene5.pdf}(a). 

For  $\theta=80^\circ$ as shown in Fig.\ \ref{populations_silicene_80_deg.pdf}(c), the $x$ component of the maximum field, $F_x<0$, promotes transfer of electrons from left to right (in the direction $x>0$) according to their negative charge -- cf.\ Fig.\ \ref{silicene5.pdf}(d). The distance of transfer is the same as in the case of the $yz$-polarized field $L_{xy}=a/\left(2\sqrt{3}\right)\approx 0.7~\mathrm{\AA}$ and adiabaticity is violated since $\delta=\hbar\omega/\left(eF_y L_{xy}\right)\gtrsim1$. Correspondingly, the residual CB populations   ${\cal N}_c^{\mathrm(res)}$ are again close to their corresponding maxima during the pulse. 

Dramatically different behavior takes place for the reciprocal incidence,  $\theta=-80^\circ$ where the CB population kinetics is displayed in Fig.\ \ref{populations_silicene_80_deg.pdf}(d). For relatively weak fields, $F_0=0.5-1.5~\mathrm{V/\AA}$, the kinetics is essentially irreversible, where the maximum CB population is attained at the end of the excitation pulse, similar to the case of Fig.\ \ref{populations_silicene_80_deg.pdf}(c) considered above in the previous paragraph. In a sharp contrast, for stronger fields,  $F_0=2-3~\mathrm{V/\AA}$, there is partial reversibility: at the end of the pulse the CB population is reduced by a factor of $\approx 2$ with respect to its maximum. This is related to improved adiabaticity, i.e.,  decreased adiabaticity parameter, $\delta=\hbar\omega/\left(eF_y L_{xx}\right)\lesssim1$ where $L_{xx}=a/\sqrt{3}\approx 1.4~\mathrm{\AA}$ is the horizontal transfer distance, see Fig.\ \ref{silicene2.pdf}(a). This distance is twice longer than for the case of Figs.\ \ref{silicene5.pdf}(c)-(d) corresponding to the polarizations in Figs.\ \ref{populations_silicene_80_deg.pdf}(b)-(c).

Note that the adiabaticity in the case of Fig.\ \ref{populations_silicene_80_deg.pdf}(d) is incomplete; for comparison, in the case of silica (quartz) a nearly perfect adiabaticity has been predicted and observed \cite{Apalkov_Stockman_PRB_2012_Strong_Field_Reflection, Schultze_et_al_Nature_2012_Controlling_Dielectrics}. This high degree of adiabaticity is most certainly related to a wide band gap, $\Delta_g$ (see also Ref.\ \cite{Krausz_Stockman_Nature_Photonics_2014_Attosecond_Review}) and to a significantly larger lattice constant, $a\approx 5~\mathrm{\AA}$, in quartz. Both these factors determine adiabaticity, which is pronounced when $\hbar\omega/\Delta_g\ll1$ and $\hbar\omega/(eF_0 a)\ll1$. Thus one should not expect near-perfect adiabaticity in graphene (cf. Ref.\ \cite{Stockman_et_al_PRB_2015_Graphene_in_Strong_Field}), silicene, and germanene where $\Delta_g$ is negligible, and $a$ is relatively small.

\begin{figure}
\begin{center}\includegraphics[width=0.48\textwidth]{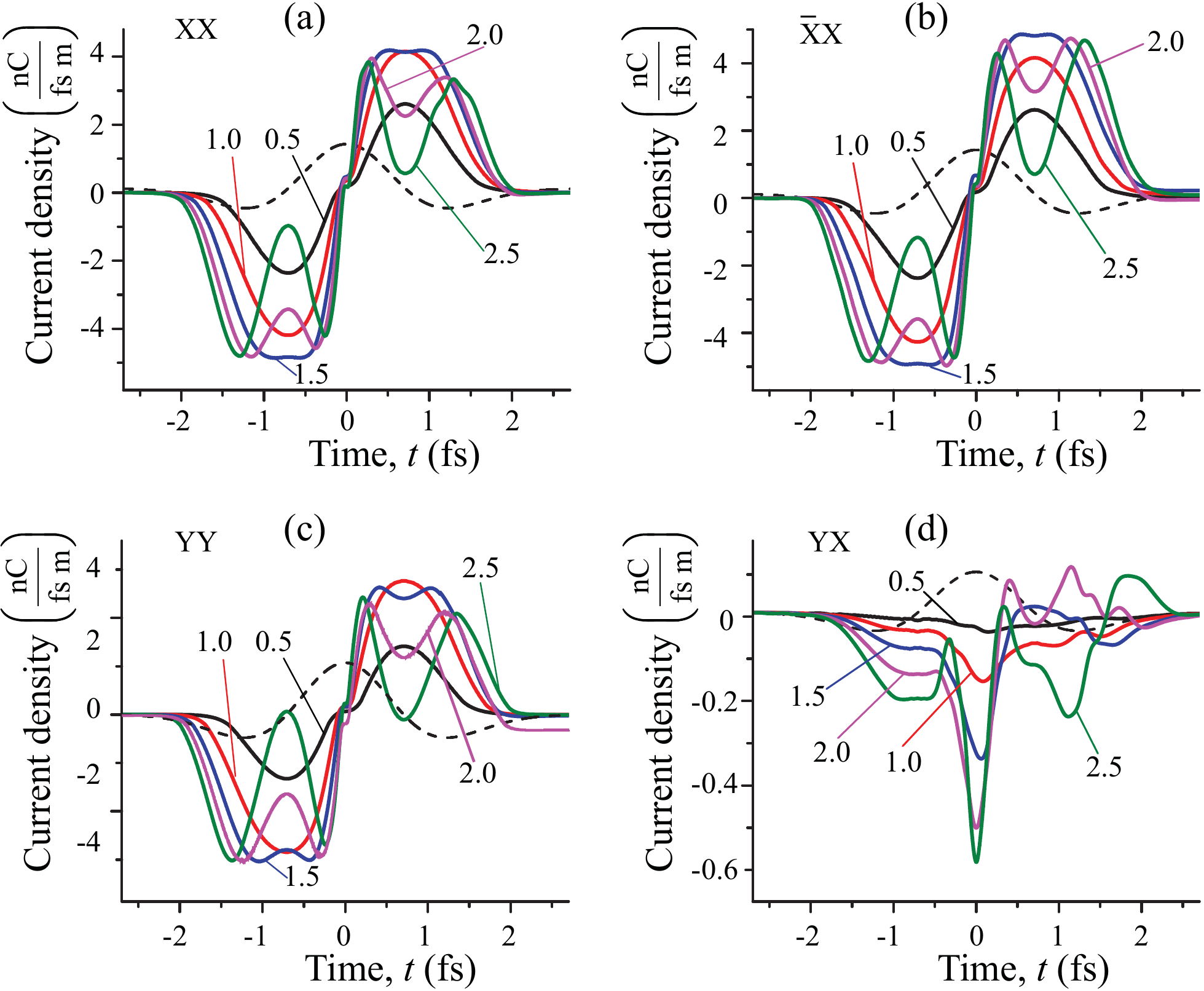}\end{center}
\caption{Current dynamics in silicene subjected to strong field pulse. The broken line displays the shape of the excitation pulse. The numbers labeling the curves are the corresponding field amplitudes $F_0~ \mathrm{(V/\AA)}$. (a) The $x$-component of the current density, $J_x$, as a function of time $t$ for the excitation pulse polarized in the $xz$ plane with the direction of the maximum in-plane field, $F_{0x}$, in the negative $x$ direction [same as in Fig.\ \ref{populations_silicene_80_deg.pdf}(c)]. (b) Same as in panel (a) but for the opposite direction of  $F_{0x}$ [as denoted as $\bar{X}X$, corresponding to Fig.\ \ref{populations_silicene_80_deg.pdf}(d)]. (c) The $y$-component of the current density, $J_y$, as a function of time $t$ for the excitation pulse polarized in the $yz$ plane. (d) Same as in panel (c) but for the $x$ component of the current density. 
} 
\label{Current_Dynamics.pdf}
\end{figure}

\subsection{Ultrafast Currents Induced by Strong Pulse}
\label{Currents}

Electrical current is due to displacement of charges caused by the applied pulse field. For free classical electrons, this current is proportional to their mean velocity, i.e., to the integral of the field, often referred to as vector potential, 
\begin{equation}
\mathbf{A}(t)=-c\int_{-\infty}^t \mathbf{F}_{2d}(t^\prime)dt^\prime~. 
\label{A}
\end{equation}
In contrast to free electrons, as we have argued above in Sec.\ \ref{Populations}, the strong field acting on the electrons in crystal lattice of silicene causes effective symmetry reduction from honeycomb to triangular and, in particular, dependence of the electron dynamics on the sign of the maximum field -- cf. Figs.\ \ref{populations_silicene_80_deg.pdf} (c) and (d). The observed partial adiabaticity is also due to the presence of the periodic lattice and defined by its period in the field direction.

The effective reduction of symmetry to triangular (where there is no inversion center)  caused by the strong normal ($z$) field component causes the currents in the silicene lattice to be highly anisotropic and non-reciprocal as we show below in this Section. Let us denote $J_{XX}$ an $x$-component of the current density induced by the field polarized in the $xz$ plane with the maximum in the negative $x$ direction as shown in  Fig.\ \ref{populations_silicene_80_deg.pdf}(c). Similarly, we denote $J_{\bar XX}$ the $x$ component of the current density caused by the field with the maximum in the positive $x$ direction as in the case of Fig.\ \ref{populations_silicene_80_deg.pdf}(d). Note that generally $J_{XX}\ne -J_{\bar XX}$ (as would have been the case for free electrons) due to the low, triangular effective symmetry. 

Similarly, we introduce current density $J_{YY}$ as the $y$ component of the current density induced by the $yz$ polarized pulse. Note that in this case, the presence of the $xz$-symmetry plane dictates that $J_{YY}=-J_{\bar YY}$. Interestingly enough, the in-plane field in the $y$ direction causes also a current in the $x$ direction [cf.\ Figs.\ \ref{silicene5.pdf} (b) and (c)], whose density we will denote as $J_{YX}$. Note that due to the symmetry,  this current is invariant with respect to inversion in the $xz$ plane, i.e., $J_{YX}=J_{\bar YX}$.

In Fig.\ \ref{Current_Dynamics.pdf}, we plot the temporal behavior of the current density for the four independent cases of the pulse polarization and current direction, $XX$, $\bar XX$, $YY$, and $YX$, as indicated in the panels; the currents in all other cases are either related to these cases by symmetry, as presented in the previous two paragraphs, or equal zero as, e.g., $J_{XY}$ and $J_{\bar XY}$. For the $XX$ case shown in Fig.\ \ref{Current_Dynamics.pdf}(a), in the relatively weak fields, $F_0\le 1~\mathrm{V/\AA}$, the current density, $J_{XX}$, obviously, qualitatively follows the vector potential, $A(t)$ reaching (negative)  maximum at approximately quarter oscillation period and turning to zero at the maximum field ($t=0$). Kinetics $J_{XX}(t)$ is approximately antisymmetric with respect to point $t=0$, which shows that this process is nearly time-reversible.

\begin{figure}
\begin{center}\includegraphics[width=0.48\textwidth]{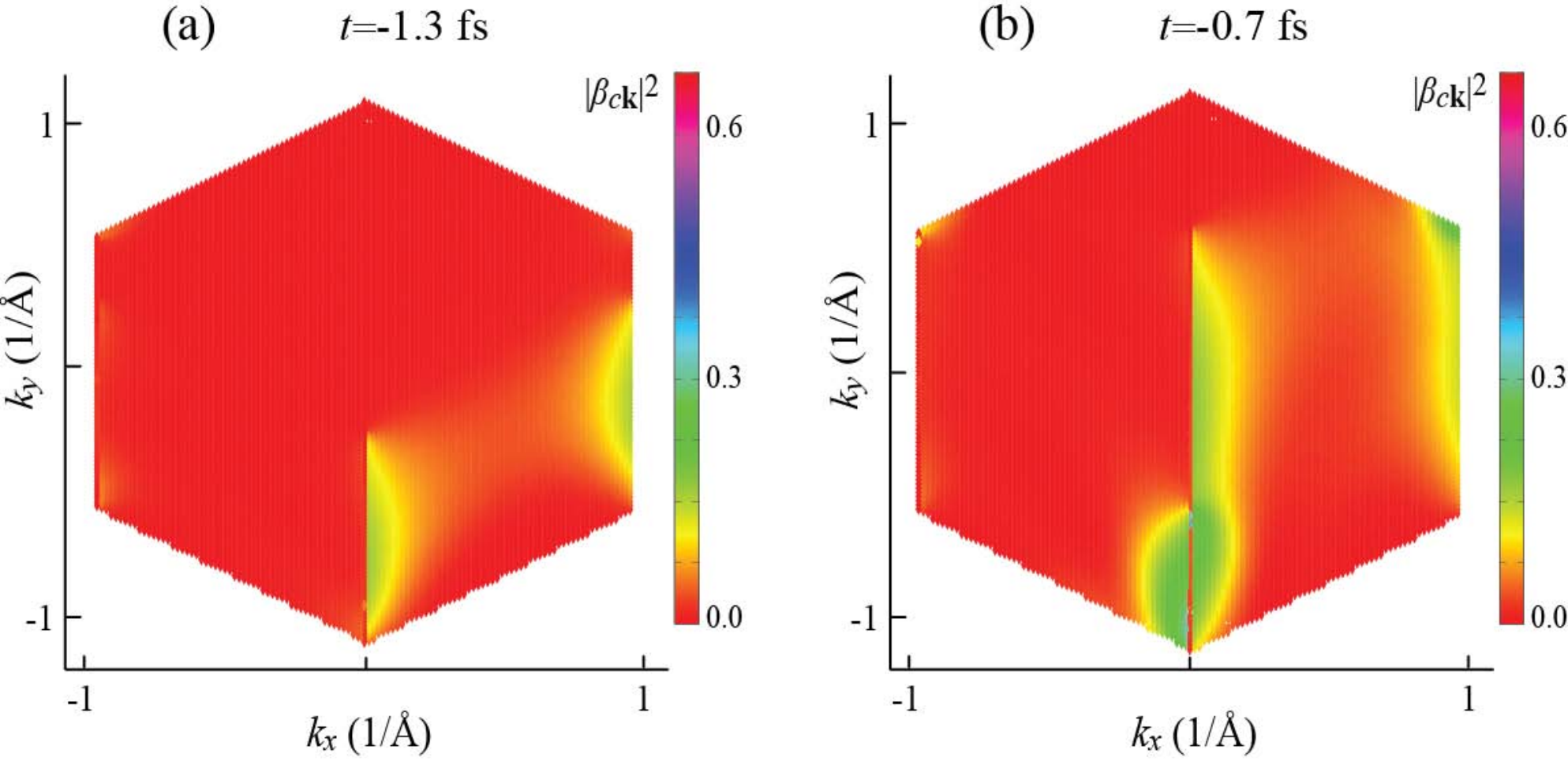}\end{center}
\caption{Electron momentum distribution $\left\vert \beta_{c\mathbf{k}}\right\vert^2$ in the first Brillouin zone in the CB for pulse with maximum field $F_0=2.5~\mathrm{V/\AA}$ with the Y-polarization of the in-plane field (the maximum in-plane field directed along the positive $y$ axis). (a) Distribution at the moment of time $t=-1.3~\mathrm{fs}$ corresponding to the maximum negative current $J_{YY}$. (b) Distribution at the moment of time $t=-0.7~\mathrm{fs}$ corresponding to the maximum positive oscillation of current $J_{YY}$.
} 
\label{kxy2.pdf}
\end{figure}

However, at higher fields, the behavior in Fig.\ \ref{Current_Dynamics.pdf}(a)  becomes nontrivial. The first manifestation of this behavior appears at  $F_0= 1.5~\mathrm{V/\AA}$ where instead of a pronounced minimum (maximum negative current) there is a plateau, which turns to a maximum for $F_0\ge 2~\mathrm{V/\AA}$. We attribute this behavior to electrons that are compelled by the field force to drift in the reciprocal space across the K-point.  We will discuss this behavior in more detail in conjunction with Fig.\ \ref{Current_Dynamics.pdf}(c) -- see below.

A phenomenon of fundamental importance is the loss of adiabaticity in higher fields, which manifests itself in the lack of anti-symmetry with respect to point $t=0$ in Fig.\ \ref{Current_Dynamics.pdf}(a). Note that non-adiabaticity also implies irreversibility
\footnote{Non-adiabaticity implies increase of entropy from the statistical or thermodynamic standpoint. Hence, non-adiabatic processes are irreversible. Examples of irreversible processes are seen in Fig.\ \ref{populations_silicene_80_deg.pdf}(b)-(c), while panel (d) shows partially reversible processes.}
 and, consequently, violation of time-reversal symmetry (called also T-invariance or T-symmetry). This violation of adiabaticity is related to a gradual transfer of population between the A and B sublattices, as we discussed above in Sec.\ \ref{Populations}. Such transfer is not instantaneous; one can estimate characteristic time it requires as $t_{tr}\sim\pi\hbar/(eL_z F_0)$. For a high field used, $F_0\sim2~\mathrm{V/\AA}$, we obtain $t_{tr}\sim 1~\mathrm{fs}$. This is in a full qualitative agreement with the results of  Fig.\ \ref{Current_Dynamics.pdf}(a) where the time-reversal  asymmetry becomes pronounced for high fields and times longer than $\sim1~\mathrm{fs}$ from the moment the pulse is applied. 

\begin{figure}
\begin{center}\includegraphics[width=0.3\textwidth]{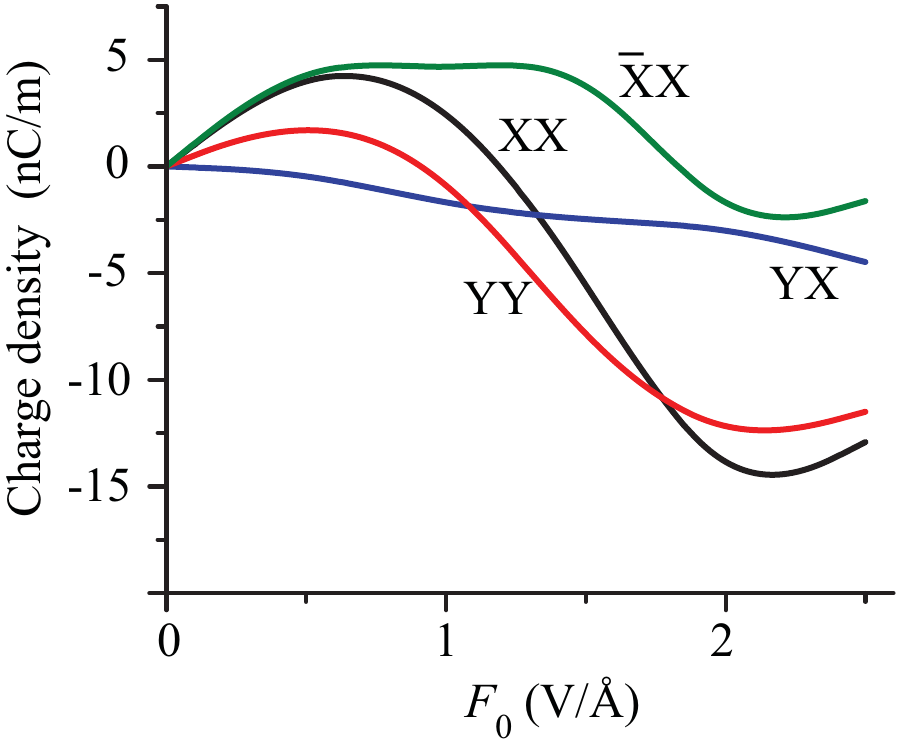}\end{center}
\caption{Charge density $Q$ transferred in the plane of the silicene layer as a function of the maximum pulse field, $F_0$. Four cases are displayed: $Q_{XX}$, $Q_{\bar XX}$, $Q_{YY}$, and $Q_{YX}$, as labeled in the Figure. All other components of the charge density transferred are either zero, or related to these four cases.
} 
\label{Charge_vs_Fnew.pdf}
\end{figure}

One of the consequences of the T-invariance violation are non-zero values of the transferred charge and of the residual polarization -- see  Eq.\ (\ref{Qtr}) and Fig.\ \ref{Charge_vs_Fnew.pdf} and the corresponding discussion -- violating the T-symmetry and adiabaticity. This implies that the system's dynamics is irreversible (non-adiabatic), which may surprise one because the system is completely Hamiltonian. This is due to the fact that the central frequency of the laser radiation, $\hbar\omega\approx 1.5~\mathrm{eV}$, is close to the transition frequency between the electron states localized at the two sublattices, $\hbar\Delta\omega\sim\pi\hbar/t_{tr}=eL_z F_0\sim 1.4~\mathrm{eV}$. This causes resonant absorption leading to dephasing -- collisionless relaxation widely known as Landau damping \cite{Landau_J_Phys_1946_Landau_Damping}.

Current kinetics for the $\bar XX$ case displayed in Fig.\ \ref{Current_Dynamics.pdf}(b) is qualitatively similar to that for the $XX$ case discussed above in the previous three paragraphs. However, the symmetry reduction caused by the nonlinear interaction with a controlled (zero in our case) CEP causes current $J_{\bar XX}$  to differ quantitatively from $J_{XX}$, which difference is pronounced in  the second half-period ($t>0$) where the T-asymmetry of the current becomes evident. The latter is due to the non-adiabaticity, already mentioned above in the discussion of Fig.\ \ref{Current_Dynamics.pdf}(a): the transfer of the electrons between sublattices occurs during a finite period of time, $t_{tr}\sim\pi\hbar/(eL_z F_0)\sim 1~\mathrm{fs}$, comparable with half optical period in our case.

The $YY$ case illustrated in Fig.\ \ref{Current_Dynamics.pdf}(c) is not related by crystal symmetry or other invariances to the $XX$ and $\bar XX$ cases considered above. However, the kinetics of $J_{YY}$ is qualitatively similar to, though quantitatively different from, the previous two cases. Note that there is strict symmetry $J_{\bar YY}=-J_{YY}$. Here also the T-symmetry is  violated: the kinetics in the first and second half-periods is dramatically different. Note that in this case, current at the end of the pulse may not vanish, which is certainly due to absence of collisions and other interactions in the model. Note that the electron-electron collisions is the fastest interaction-induced relaxation process. However, it takes the electron-electron collisions  $\sim 10-20~\mathrm{fs}$ in a similar two-dimensional system, graphene,  to make an effect  \cite{Ultrafast_collinear_scattering_graphene_nat_comm_2013}, which is too long time for our pulse whose entire duration is less than 4 fs.

The results for current $J_{YX}$ (in the $x$ direction induced by the field in the $y$ direction) are displayed in Fig.\ \ref{Current_Dynamics.pdf}(d). Note that exactly $J_{YX}=J_{\bar YX}$ due to symmetry. Without an electric field applied, silicene is a center-symmetric solid. Therefore for low fields current $J_{YX}$ should vanish. This is, in fact, the case with a good accuracy for $F_0=0.5~\mathrm{V/\AA}$, as one can see in the Fig.\ \ref{Current_Dynamics.pdf}(d). With field increasing, there is an increased  current  $J_{YX}$. Predominantly, it is directed along the negative $x$ axis, as is understandable from comparison with Figs.\ \ref{silicene5.pdf}(b) and (c). Note that magnitude of this current is approximately an order of magnitude smaller than $J_{YY}$. 

The origin of the current oscillations for strong fields, $F_0\ge 2~\mathrm{V/\AA}$, in Figs.\ \ref{Current_Dynamics.pdf}(a)-(c) can be understood from the electron momentum distribution. Consider for certainty the YY case, where the current is shown in  Fig.\ \ref{Current_Dynamics.pdf}(c). The corresponding momentum distribution for electrons in the CB,  $\left\vert \beta_{c\mathbf{k}}\right\vert^2$, for pulse field amplitude $F_0=2.5~\mathrm{V/\AA}$ is displayed in in Fig.\ \ref{kxy2.pdf}(a) for moment of time $t=-1.3~\mathrm{fs}$, corresponding to the minimum (the maximum negative value) of the current, $J_{YY}$. At this instance, which just precedes the current oscillation, excess electron population (depicted by green) is concentrated at $k_y<0$, $k_x=0$. This excess population is formed due to field force $eF_y(t)>0$ that pushes the electrons across the K points into the second Brillouin zone in the extended zone picture; these electrons appear in the first Brillouin zone at the K point at $k_y\approx -1~\mathrm{\AA^{-1}}$, $k_x=0$. The second localization of electrons is around a K$^\prime$ point at $k_x=1~\mathrm{\AA^{-1}}$. This electron population is formed due to the lack of the center symmetry in the presence of strong field $F_z$, i.e., it has the same origin at current $J_{YX}$ described above in the previous paragraph.

Dramatically different electron distribution is displayed in  Fig.\ \ref{kxy2.pdf}(b) for  $t=-0.7~\mathrm{fs}$ when current $J_{YY}$ experiences the maximum upswing. This is caused by a significant number of electrons in the CB with $k_y>0$ which appear due to a drift in the reciprocal space under force $eF_y>0$. These electrons make a positive contribution to the current (their group velocity $v_g<0$; correspondingly, due to $e<0$, their contribution to $J_{YY}$ is positive). 

The currents described above in conjunction with Fig.\ \ref{Current_Dynamics.pdf} cause transfer of charge across the system and accumulation of charges by the end of the pulse as given by Eq.\ (\ref{Qtr}). Such charge $Q$ transferred through the system is displayed as a function of the field amplitude, $F_0$, in Fig.\ \ref{Charge_vs_Fnew.pdf} for four independent combinations of the field and current directions, $XX$, $\bar XX$, $YY$, and $YX$. A remarkable property of these results is that in all cases, except for $YX$, the charge transferred changes its sign as the field amplitude increases. This can be attributed to the increased number of electrons experiencing the Bragg reflections at the Brillouin zone boundary, especially at the K-points, with the field increase. Thus this sign change of the transferred charge has the same origin as the current oscillations in  Fig.\ \ref{Current_Dynamics.pdf} as described above. This charge accumulated at the pulse end is an experimentally observable quantity just as the previous experiments on currents in dielectrics \cite{Schiffrin_at_al_Nature_2012_Current_in_Dielectric, Stockman_et_al_Nat_Phot_2013_CEP_Detector}. On the order of magnitude, this accumulated charge  in Fig.\ \ref{Charge_vs_Fnew.pdf} is $Q\sim 1~\mathrm{fC/\mu m}$. For a $\sim 1~\mathrm{\mu m}$ focused spot, this gives a $\sim 1~\mathrm{fC}$ transferred charge. Such a charge is on the same order of magnitude as in experiments Refs.\ \cite{Schiffrin_at_al_Nature_2012_Current_in_Dielectric, Stockman_et_al_Nat_Phot_2013_CEP_Detector} and is, in principle, reliably observable.

\section{Concluding Discussion}
\label{Conclusion}

The accumulation of charge $Q$ transferred through the system implies a dramatic manifestation of  symmetry violation. This  charge accumulation violates simultaneously the parity symmetry (P-symmetry) and the charge-inversion symmetry (C-symmetry). This violation happens due to the fact that our pulse is short and has a controlled CEP (zero in our case): the maximum field is reached at the maximum of the envelope (instance $t=0$). Due to strong nonlinearity of the system for fields $F_0\gtrsim 1~\mathrm{V/\AA}$ applied,  this maximum field defines a selected direction in the system plane for the force acting on electrons. This causes the violation of the C- and P-symmetries. This is actually a general property for systems subjected to a short, strong, CEP-controlled pulses. It takes place in both two-dimensional solids such as graphene, silicene, germanene, and also conventional three-dimensional solids such as fused silica, sapphire, etc. In particular, it was a fundamental origin for the charge transfer in silica and quartz in original experiments \cite{Schiffrin_at_al_Nature_2012_Current_in_Dielectric}.

A symmetry violation specific for silicene is related to the electron transfer between the sublattices caused by the normal field component $F_z$, which effectively reduces the system's symmetry from hexagonal to triangular. This causes non-reciprocity: $J_{\bar XX}\ne J_{XX}$ and the appearance of a cross-current, $J_{XY}\ne 0$. Note that anisotropicity in the $xy$ plane, $J_{XX}\ne J_{YY}$, is inherent in both silicene and graphene.

Our zero-CEP pulse is T-symmetric; in the absence of the T-symmetry violation, the current should be T-odd, which would preclude the accumulation of charges after the pulse. However, we have seen from the results of  Figs.\ \ref{Current_Dynamics.pdf}(a)-(c) that the current is not anti-symmetric in time, i.e., there is a significant violation of the T-symmetry, which we attribute to the Landau damping. This is inherent in both graphene and silicene and is due to the absence of a significant band gap; this is in contrast to silica that is almost perfectly T-reversible \cite{Schiffrin_at_al_Nature_2012_Current_in_Dielectric, Schultze_et_al_Nature_2012_Controlling_Dielectrics}. An additional contribution to T-irreversibility stems from the fact that the frequency associated with the electron transfer in the normal direction, $eF_z L_z/\hbar$, is on the same order as the carrier frequency of the pulse. This causes resonant absorption of the excitation pulse and the Landau damping, specific for the silicene (and also germanene). If adiabaticity were present, it would have guaranteed reversibility and would have forbidden the charge accumulation. 

Finally, we note a close analogy of silicene with the field-effect transistor (FET) \cite{Kahng_MOSFET_Patent_1963, Taur_Ning_1998_VLSI_Fundamentals, Liou_Schwierz_2003_Microwave_Transistors, Schwierz_Wong_Liou_2010_Nanometer_CMOS}. In FET, the gate field, applied normally to the conducting channel, changes the carrier populations in it and, thereby, controls its conductance. Analogously, in silicene, the normal field component, $F_z$, transfers carriers to one of the sublattices, A or B, thereby changing the system's response to the in-plane field. A fundamental difference (and advantage) of silicene is that such a ``device'' works at optical frequencies, with the response time on the (sub)femtosecond scale. This opens a potential for many applications of silicene in future Petahertz-speed devices and applications. 

\section*{Acknowledgment}

The work of VA was supported by Grant No. ECCS-1308473 from NSF. This work of MIS and HKK was supported by major funding from  from MURI Grant FA9550-15-1-0037 from the U.S. Air Force Office of Scientific Research;  supplementary funding came from Grant No. DE-FG02-11ER46789 from the Materials Sciences and Engineering Division of the Office of the Basic Energy Sciences, Office of Science, U.S. Department of Energy and Grant No. DE-FG02-01ER15213 from the Chemical Sciences, Biosciences and Geosciences Division of the Office of the Basic Energy Sciences, Office of Science, U.S. Department of Energy.


\begin{thebibliography}{52}
\expandafter\ifx\csname natexlab\endcsname\relax\def\natexlab#1{#1}\fi
\expandafter\ifx\csname bibnamefont\endcsname\relax
  \def\bibnamefont#1{#1}\fi
\expandafter\ifx\csname bibfnamefont\endcsname\relax
  \def\bibfnamefont#1{#1}\fi
\expandafter\ifx\csname citenamefont\endcsname\relax
  \def\citenamefont#1{#1}\fi
\expandafter\ifx\csname url\endcsname\relax
  \def\url#1{\texttt{#1}}\fi
\expandafter\ifx\csname urlprefix\endcsname\relax\def\urlprefix{URL }\fi
\providecommand{\bibinfo}[2]{#2}
\providecommand{\eprint}[2][]{\url{#2}}

\bibitem[{\citenamefont{Kara et~al.}(2012)\citenamefont{Kara, Enriquez,
  Seitsonen, Lew Yan~Voon, Vizzini, Aufray, and
  Oughaddou}}]{review_silicene_2012}
\bibinfo{author}{\bibfnamefont{A.}~\bibnamefont{Kara}},
  \bibinfo{author}{\bibfnamefont{H.}~\bibnamefont{Enriquez}},
  \bibinfo{author}{\bibfnamefont{A.~P.} \bibnamefont{Seitsonen}},
  \bibinfo{author}{\bibfnamefont{L.~C.} \bibnamefont{Lew Yan~Voon}},
  \bibinfo{author}{\bibfnamefont{S.}~\bibnamefont{Vizzini}},
  \bibinfo{author}{\bibfnamefont{B.}~\bibnamefont{Aufray}}, \bibnamefont{and}
  \bibinfo{author}{\bibfnamefont{H.}~\bibnamefont{Oughaddou}},
  \bibinfo{journal}{Surf. Sci. Rep.} \textbf{\bibinfo{volume}{67}},
  \bibinfo{pages}{1} (\bibinfo{year}{2012}).

\bibitem[{\citenamefont{Tang and
  Zhou}(2013)}]{graphene_analogous_materials_2013}
\bibinfo{author}{\bibfnamefont{Q.}~\bibnamefont{Tang}} \bibnamefont{and}
  \bibinfo{author}{\bibfnamefont{Z.}~\bibnamefont{Zhou}},
  \bibinfo{journal}{Progr. Mater. Sci.} \textbf{\bibinfo{volume}{58}},
  \bibinfo{pages}{1244} (\bibinfo{year}{2013}).

\bibitem[{\citenamefont{Rostami et~al.}(2013)\citenamefont{Rostami, Moghaddam,
  and Asgari}}]{Effective_lattice_Hamiltonian_MoS2_PRB_2013}
\bibinfo{author}{\bibfnamefont{H.}~\bibnamefont{Rostami}},
  \bibinfo{author}{\bibfnamefont{A.~G.}~\bibnamefont{Moghaddam}},
  \bibnamefont{and} \bibinfo{author}{\bibfnamefont{R.}~\bibnamefont{Asgari}},
  \bibinfo{journal}{Phys. Rev. B} \textbf{\bibinfo{volume}{88}},
  \bibinfo{pages}{085440} (\bibinfo{year}{2013}).

\bibitem[{\citenamefont{Kormanyos et~al.}(2013)\citenamefont{Kormanyos,
  Zolyomi, Drummond, Rakyta, Burkard, and Fal'ko}}]{Monolayer_MoS2_PRB_2013}
\bibinfo{author}{\bibfnamefont{A.}~\bibnamefont{Kormanyos}},
  \bibinfo{author}{\bibfnamefont{V.}~\bibnamefont{Zolyomi}},
  \bibinfo{author}{\bibfnamefont{N.~D.}~\bibnamefont{Drummond}},
  \bibinfo{author}{\bibfnamefont{P.}~\bibnamefont{Rakyta}},
  \bibinfo{author}{\bibfnamefont{G.}~\bibnamefont{Burkard}}, \bibnamefont{and}
  \bibinfo{author}{\bibfnamefont{V.~I.}~\bibnamefont{Fal'ko}},
  \bibinfo{journal}{Phys. Rev. B} \textbf{\bibinfo{volume}{88}},
  \bibinfo{pages}{045416} (\bibinfo{year}{2013}).

\bibitem[{\citenamefont{Takeda and
  Shiraishi}(1994)}]{corrugation_Si_Ge_PRB_1994}
\bibinfo{author}{\bibfnamefont{K.}~\bibnamefont{Takeda}} \bibnamefont{and}
  \bibinfo{author}{\bibfnamefont{K.}~\bibnamefont{Shiraishi}},
  \bibinfo{journal}{Phys. Rev. B} \textbf{\bibinfo{volume}{50}},
  \bibinfo{pages}{14916} (\bibinfo{year}{1994}).

\bibitem[{\citenamefont{Liu et~al.}(2011{\natexlab{a}})\citenamefont{Liu, Feng,
  and Yao}}]{Quantum_Spin_Hall_effect_PRL_2011}
\bibinfo{author}{\bibfnamefont{C.-C.} \bibnamefont{Liu}},
  \bibinfo{author}{\bibfnamefont{W.}~\bibnamefont{Feng}}, \bibnamefont{and}
  \bibinfo{author}{\bibfnamefont{Y.}~\bibnamefont{Yao}},
  \bibinfo{journal}{Phys. Rev. Lett.} \textbf{\bibinfo{volume}{107}},
  \bibinfo{pages}{076802} (\bibinfo{year}{2011}{\natexlab{a}}).

\bibitem[{\citenamefont{Liu et~al.}(2011{\natexlab{b}})\citenamefont{Liu,
  Jiang, and Yao}}]{Low_energy_Hamiltonian_silicene_germanene_PRB_2011}
\bibinfo{author}{\bibfnamefont{C.-C.} \bibnamefont{Liu}},
  \bibinfo{author}{\bibfnamefont{H.}~\bibnamefont{Jiang}}, \bibnamefont{and}
  \bibinfo{author}{\bibfnamefont{Y.}~\bibnamefont{Yao}},
  \bibinfo{journal}{Phys. Rev. B} \textbf{\bibinfo{volume}{84}},
  \bibinfo{pages}{195430} (\bibinfo{year}{2011}{\natexlab{b}}).

\bibitem[{\citenamefont{Wang et~al.}(2012)\citenamefont{Wang, Zheng, Ni, Fei,
  Liu, Quhe, Xu, Zhou, Gao, and
  Lu}}]{Half_metallic_silicene_germanene_Nano_2012}
\bibinfo{author}{\bibfnamefont{Y.}~\bibnamefont{Wang}},
  \bibinfo{author}{\bibfnamefont{J.}~\bibnamefont{Zheng}},
  \bibinfo{author}{\bibfnamefont{Z.}~\bibnamefont{Ni}},
  \bibinfo{author}{\bibfnamefont{R.}~\bibnamefont{Fei}},
  \bibinfo{author}{\bibfnamefont{Q.}~\bibnamefont{Liu}},
  \bibinfo{author}{\bibfnamefont{R.}~\bibnamefont{Quhe}},
  \bibinfo{author}{\bibfnamefont{C.}~\bibnamefont{Xu}},
  \bibinfo{author}{\bibfnamefont{J.}~\bibnamefont{Zhou}},
  \bibinfo{author}{\bibfnamefont{Z.}~\bibnamefont{Gao}}, \bibnamefont{and}
  \bibinfo{author}{\bibfnamefont{J.}~\bibnamefont{Lu}}, \bibinfo{journal}{Nano}
  \textbf{\bibinfo{volume}{07}}, \bibinfo{pages}{1250037}
  (\bibinfo{year}{2012}).

\bibitem[{\citenamefont{Zheng and
  Zhang}(2012)}]{Functionalized_silicene_Nano_research_2012}
\bibinfo{author}{\bibfnamefont{F.-b.} \bibnamefont{Zheng}} \bibnamefont{and}
  \bibinfo{author}{\bibfnamefont{C.-w.} \bibnamefont{Zhang}},
  \bibinfo{journal}{Nanoscale Res. Lett.} \textbf{\bibinfo{volume}{7}},
  \bibinfo{pages}{422} (\bibinfo{year}{2012}).

\bibitem[{\citenamefont{Lalmi et~al.}(2010)\citenamefont{Lalmi, Oughaddou,
  Enriquez, Kara, Vizzini, Ealet, and Aufray}}]{silicene_APL_2010}
\bibinfo{author}{\bibfnamefont{B.}~\bibnamefont{Lalmi}},
  \bibinfo{author}{\bibfnamefont{H.}~\bibnamefont{Oughaddou}},
  \bibinfo{author}{\bibfnamefont{H.}~\bibnamefont{Enriquez}},
  \bibinfo{author}{\bibfnamefont{A.}~\bibnamefont{Kara}},
  \bibinfo{author}{\bibfnamefont{S.}~\bibnamefont{Vizzini}},
  \bibinfo{author}{\bibfnamefont{B.}~\bibnamefont{Ealet}}, \bibnamefont{and}
  \bibinfo{author}{\bibfnamefont{B.}~\bibnamefont{Aufray}},
  \bibinfo{journal}{Appl. Phys. Lett.} \textbf{\bibinfo{volume}{97}},
  \bibinfo{pages}{223109} (\bibinfo{year}{2010}).

\bibitem[{\citenamefont{De~Padova et~al.}(2011)\citenamefont{De~Padova,
  Quaresima, Olivieri, Perfetti, and Le~Lay}}]{silicene_APL_2011}
\bibinfo{author}{\bibfnamefont{P.}~\bibnamefont{De~Padova}},
  \bibinfo{author}{\bibfnamefont{C.}~\bibnamefont{Quaresima}},
  \bibinfo{author}{\bibfnamefont{B.}~\bibnamefont{Olivieri}},
  \bibinfo{author}{\bibfnamefont{P.}~\bibnamefont{Perfetti}}, \bibnamefont{and}
  \bibinfo{author}{\bibfnamefont{G.}~\bibnamefont{Le~Lay}},
  \bibinfo{journal}{Appl. Phys. Lett.} \textbf{\bibinfo{volume}{98}},
  \bibinfo{pages}{081909} (\bibinfo{year}{2011}).

\bibitem[{\citenamefont{Vogt et~al.}(2012)\citenamefont{Vogt, De~Padova,
  Quaresima, Avila, Frantzeskakis, Asensio, Resta, Ealet, and
  Le~Lay}}]{silicene_exp_PRL_2012_Vogt}
\bibinfo{author}{\bibfnamefont{P.}~\bibnamefont{Vogt}},
  \bibinfo{author}{\bibfnamefont{P.}~\bibnamefont{De~Padova}},
  \bibinfo{author}{\bibfnamefont{C.}~\bibnamefont{Quaresima}},
  \bibinfo{author}{\bibfnamefont{J.}~\bibnamefont{Avila}},
  \bibinfo{author}{\bibfnamefont{E.}~\bibnamefont{Frantzeskakis}},
  \bibinfo{author}{\bibfnamefont{M.~C.} \bibnamefont{Asensio}},
  \bibinfo{author}{\bibfnamefont{A.}~\bibnamefont{Resta}},
  \bibinfo{author}{\bibfnamefont{B.}~\bibnamefont{Ealet}}, \bibnamefont{and}
  \bibinfo{author}{\bibfnamefont{G.}~\bibnamefont{Le~Lay}},
  \bibinfo{journal}{Phys. Rev. Lett.} \textbf{\bibinfo{volume}{108}},
  \bibinfo{pages}{155501} (\bibinfo{year}{2012}).

\bibitem[{\citenamefont{Chen et~al.}(2012)\citenamefont{Chen, Liu, Feng, He,
  Cheng, Ding, Meng, Yao, and Wu}}]{Evidence_silicene_PRL_2012}
\bibinfo{author}{\bibfnamefont{L.}~\bibnamefont{Chen}},
  \bibinfo{author}{\bibfnamefont{C.-C.} \bibnamefont{Liu}},
  \bibinfo{author}{\bibfnamefont{B.}~\bibnamefont{Feng}},
  \bibinfo{author}{\bibfnamefont{X.}~\bibnamefont{He}},
  \bibinfo{author}{\bibfnamefont{P.}~\bibnamefont{Cheng}},
  \bibinfo{author}{\bibfnamefont{Z.}~\bibnamefont{Ding}},
  \bibinfo{author}{\bibfnamefont{S.}~\bibnamefont{Meng}},
  \bibinfo{author}{\bibfnamefont{Y.}~\bibnamefont{Yao}}, \bibnamefont{and}
  \bibinfo{author}{\bibfnamefont{K.}~\bibnamefont{Wu}}, \bibinfo{journal}{Phys.
  Rev. Lett.} \textbf{\bibinfo{volume}{109}}, \bibinfo{pages}{056804}
  (\bibinfo{year}{2012}).

\bibitem[{\citenamefont{Chun-Liang et~al.}(2012)\citenamefont{Chun-Liang,
  Ryuichi, Kazuaki, Noriyuki, Emi, Yousoo, Noriaki, and
  Maki}}]{Structure_silicene_applied_physics_express_2012}
\bibinfo{author}{\bibfnamefont{L.}~\bibnamefont{Chun-Liang}},
  \bibinfo{author}{\bibfnamefont{A.}~\bibnamefont{Ryuichi}},
  \bibinfo{author}{\bibfnamefont{K.}~\bibnamefont{Kazuaki}},
  \bibinfo{author}{\bibfnamefont{T.}~\bibnamefont{Noriyuki}},
  \bibinfo{author}{\bibfnamefont{M.}~\bibnamefont{Emi}},
  \bibinfo{author}{\bibfnamefont{K.}~\bibnamefont{Yousoo}},
  \bibinfo{author}{\bibfnamefont{T.}~\bibnamefont{Noriaki}}, \bibnamefont{and}
  \bibinfo{author}{\bibfnamefont{K.}~\bibnamefont{Maki}},
  \bibinfo{journal}{Appl. Phys. Expr.} \textbf{\bibinfo{volume}{5}},
  \bibinfo{pages}{045802} (\bibinfo{year}{2012}).

\bibitem[{\citenamefont{Feng et~al.}(2012)\citenamefont{Feng, Ding, Meng, Yao,
  He, Cheng, Chen, and Wu}}]{evidence_silicene_Nano_letters_2012}
\bibinfo{author}{\bibfnamefont{B.}~\bibnamefont{Feng}},
  \bibinfo{author}{\bibfnamefont{Z.}~\bibnamefont{Ding}},
  \bibinfo{author}{\bibfnamefont{S.}~\bibnamefont{Meng}},
  \bibinfo{author}{\bibfnamefont{Y.}~\bibnamefont{Yao}},
  \bibinfo{author}{\bibfnamefont{X.}~\bibnamefont{He}},
  \bibinfo{author}{\bibfnamefont{P.}~\bibnamefont{Cheng}},
  \bibinfo{author}{\bibfnamefont{L.}~\bibnamefont{Chen}}, \bibnamefont{and}
  \bibinfo{author}{\bibfnamefont{K.}~\bibnamefont{Wu}}, \bibinfo{journal}{Nano
  Lett.} \textbf{\bibinfo{volume}{12}}, \bibinfo{pages}{3507}
  (\bibinfo{year}{2012}).

\bibitem[{\citenamefont{Fleurence et~al.}(2012)\citenamefont{Fleurence,
  Friedlein, Ozaki, Kawai, Wang, and
  Yamada-Takamura}}]{Silicene_on_diboride_PRL_2012}
\bibinfo{author}{\bibfnamefont{A.}~\bibnamefont{Fleurence}},
  \bibinfo{author}{\bibfnamefont{R.}~\bibnamefont{Friedlein}},
  \bibinfo{author}{\bibfnamefont{T.}~\bibnamefont{Ozaki}},
  \bibinfo{author}{\bibfnamefont{H.}~\bibnamefont{Kawai}},
  \bibinfo{author}{\bibfnamefont{Y.}~\bibnamefont{Wang}}, \bibnamefont{and}
  \bibinfo{author}{\bibfnamefont{Y.}~\bibnamefont{Yamada-Takamura}},
  \bibinfo{journal}{Phys. Rev. Lett.} \textbf{\bibinfo{volume}{108}},
  \bibinfo{pages}{245501} (\bibinfo{year}{2012}).

\bibitem[{\citenamefont{DÃ¡vila et~al.}(2014)\citenamefont{DÃ¡vila, Xian,
  Cahangirov, Rubio, and Lay}}]{germanene_experimental_New_J_Physics_2014}
\bibinfo{author}{\bibfnamefont{M.~E.} \bibnamefont{DÃ¡vila}},
  \bibinfo{author}{\bibfnamefont{L.}~\bibnamefont{Xian}},
  \bibinfo{author}{\bibfnamefont{S.}~\bibnamefont{Cahangirov}},
  \bibinfo{author}{\bibfnamefont{A.}~\bibnamefont{Rubio}}, \bibnamefont{and}
  \bibinfo{author}{\bibfnamefont{G.~L.} \bibnamefont{Lay}},
  \bibinfo{journal}{New J. Phys.} \textbf{\bibinfo{volume}{16}},
  \bibinfo{pages}{095002} (\bibinfo{year}{2014}).

\bibitem[{\citenamefont{Li et~al.}(2014)\citenamefont{Li, Lu, Pan, Qin, Wang,
  Wang, Cao, Du, and Gao}}]{Germanene_on_Pt_advanced_materials_2014}
\bibinfo{author}{\bibfnamefont{L.}~\bibnamefont{Li}},
  \bibinfo{author}{\bibfnamefont{S.-z.} \bibnamefont{Lu}},
  \bibinfo{author}{\bibfnamefont{J.}~\bibnamefont{Pan}},
  \bibinfo{author}{\bibfnamefont{Z.}~\bibnamefont{Qin}},
  \bibinfo{author}{\bibfnamefont{Y.-q.} \bibnamefont{Wang}},
  \bibinfo{author}{\bibfnamefont{Y.}~\bibnamefont{Wang}},
  \bibinfo{author}{\bibfnamefont{G.-y.} \bibnamefont{Cao}},
  \bibinfo{author}{\bibfnamefont{S.}~\bibnamefont{Du}}, \bibnamefont{and}
  \bibinfo{author}{\bibfnamefont{H.-J.} \bibnamefont{Gao}},
  \bibinfo{journal}{Adv. Mat.} \textbf{\bibinfo{volume}{26}},
  \bibinfo{pages}{4820} (\bibinfo{year}{2014}).

\bibitem[{\citenamefont{Tao et~al.}(2015)\citenamefont{Tao, Cinquanta, Chiappe,
  Grazianetti, Fanciulli, Dubey, Molle, and
  Akinwande}}]{Tao_et_al_nnano.2014.325_2015_Silicine_FET}
\bibinfo{author}{\bibfnamefont{L.}~\bibnamefont{Tao}},
  \bibinfo{author}{\bibfnamefont{E.}~\bibnamefont{Cinquanta}},
  \bibinfo{author}{\bibfnamefont{D.}~\bibnamefont{Chiappe}},
  \bibinfo{author}{\bibfnamefont{C.}~\bibnamefont{Grazianetti}},
  \bibinfo{author}{\bibfnamefont{M.}~\bibnamefont{Fanciulli}},
  \bibinfo{author}{\bibfnamefont{M.}~\bibnamefont{Dubey}},
  \bibinfo{author}{\bibfnamefont{A.}~\bibnamefont{Molle}}, \bibnamefont{and}
  \bibinfo{author}{\bibfnamefont{D.}~\bibnamefont{Akinwande}},
  \bibinfo{journal}{Nat. Nano} \textbf{\bibinfo{volume}{doi:
  10.1038/nnano.2014.325}} (\bibinfo{year}{2015}).

\bibitem[{\citenamefont{Kahng}(1963)}]{Kahng_MOSFET_Patent_1963}
\bibinfo{author}{\bibfnamefont{D.}~\bibnamefont{Kahng}},
  \bibinfo{journal}{United States Patent 3,102,230}  (\bibinfo{year}{1963}).

\bibitem[{\citenamefont{Taur and
  Ning}(1998)}]{Taur_Ning_1998_VLSI_Fundamentals}
\bibinfo{author}{\bibfnamefont{Y.}~\bibnamefont{Taur}} \bibnamefont{and}
  \bibinfo{author}{\bibfnamefont{T.~H.} \bibnamefont{Ning}},
  \emph{\bibinfo{title}{Fundamentals of Modern VLSI Devices}}
  (\bibinfo{publisher}{Cambridge University Press}, \bibinfo{year}{1998}).

\bibitem[{\citenamefont{Liou and
  Schwierz}(2003)}]{Liou_Schwierz_2003_Microwave_Transistors}
\bibinfo{author}{\bibfnamefont{J.~J.} \bibnamefont{Liou}} \bibnamefont{and}
  \bibinfo{author}{\bibfnamefont{F.}~\bibnamefont{Schwierz}},
  \emph{\bibinfo{title}{Modern Microwave Transistors: Theory, Design and
  Performance}} (\bibinfo{publisher}{Wiley-Interscience},
  \bibinfo{year}{2003}).

\bibitem[{\citenamefont{Schwierz et~al.}(2010)\citenamefont{Schwierz, Wong, and
  Liou}}]{Schwierz_Wong_Liou_2010_Nanometer_CMOS}
\bibinfo{author}{\bibfnamefont{F.}~\bibnamefont{Schwierz}},
  \bibinfo{author}{\bibfnamefont{H.}~\bibnamefont{Wong}}, \bibnamefont{and}
  \bibinfo{author}{\bibfnamefont{J.~J.} \bibnamefont{Liou}},
  \emph{\bibinfo{title}{Nanometer CMOS}} (\bibinfo{publisher}{Pan Stanford},
  \bibinfo{year}{2010}).

\bibitem[{\citenamefont{Meng et~al.}(2013)\citenamefont{Meng, Wang, Zhang, Du,
  Wu, Li, Zhang, Li, Zhou, Hofer et~al.}}]{buckled_silicene_on_Ir_Nano_2012}
\bibinfo{author}{\bibfnamefont{L.}~\bibnamefont{Meng}},
  \bibinfo{author}{\bibfnamefont{Y.}~\bibnamefont{Wang}},
  \bibinfo{author}{\bibfnamefont{L.}~\bibnamefont{Zhang}},
  \bibinfo{author}{\bibfnamefont{S.}~\bibnamefont{Du}},
  \bibinfo{author}{\bibfnamefont{R.}~\bibnamefont{Wu}},
  \bibinfo{author}{\bibfnamefont{L.}~\bibnamefont{Li}},
  \bibinfo{author}{\bibfnamefont{Y.}~\bibnamefont{Zhang}},
  \bibinfo{author}{\bibfnamefont{G.}~\bibnamefont{Li}},
  \bibinfo{author}{\bibfnamefont{H.}~\bibnamefont{Zhou}},
  \bibinfo{author}{\bibfnamefont{W.~A.} \bibnamefont{Hofer}},
  \bibnamefont{et~al.}, \bibinfo{journal}{Nano Lett.}
  \textbf{\bibinfo{volume}{13}}, \bibinfo{pages}{685} (\bibinfo{year}{2013}).

\bibitem[{\citenamefont{Liu et~al.}(2011{\natexlab{c}})\citenamefont{Liu,
  Jiang, and Yao}}]{Silicene_effective_Hamiltonian_PRB_2011}
\bibinfo{author}{\bibfnamefont{C.-C.} \bibnamefont{Liu}},
  \bibinfo{author}{\bibfnamefont{H.}~\bibnamefont{Jiang}}, \bibnamefont{and}
  \bibinfo{author}{\bibfnamefont{Y.}~\bibnamefont{Yao}},
  \bibinfo{journal}{Phys. Rev. B} \textbf{\bibinfo{volume}{84}},
  \bibinfo{pages}{195430} (\bibinfo{year}{2011}{\natexlab{c}}).

\bibitem[{\citenamefont{Gmitra et~al.}(2009)\citenamefont{Gmitra, Konschuh,
  Ertler, Ambrosch-Draxl, and
  Fabian}}]{Band_structure_topologies_graphene_PRB_2009}
\bibinfo{author}{\bibfnamefont{M.}~\bibnamefont{Gmitra}},
  \bibinfo{author}{\bibfnamefont{S.}~\bibnamefont{Konschuh}},
  \bibinfo{author}{\bibfnamefont{C.}~\bibnamefont{Ertler}},
  \bibinfo{author}{\bibfnamefont{C.}~\bibnamefont{Ambrosch-Draxl}},
  \bibnamefont{and} \bibinfo{author}{\bibfnamefont{J.}~\bibnamefont{Fabian}},
  \bibinfo{journal}{Phys. Rev. B} \textbf{\bibinfo{volume}{80}},
  \bibinfo{pages}{235431} (\bibinfo{year}{2009}).

\bibitem[{\citenamefont{Ezawa}(2012)}]{silicene_inhomogeneous_field_New_J_phys%
_2012}
\bibinfo{author}{\bibfnamefont{M.}~\bibnamefont{Ezawa}}, \bibinfo{journal}{New
  J. Phys.} \textbf{\bibinfo{volume}{14}}, \bibinfo{pages}{033003}
  (\bibinfo{year}{2012}).

\bibitem[{\citenamefont{Schiffrin et~al.}(2012)\citenamefont{Schiffrin,
  Paasch-Colberg, Karpowicz, Apalkov, Gerster, Muhlbrandt, Korbman, Reichert,
  Schultze, Holzner
  et~al.}}]{Schiffrin_at_al_Nature_2012_Current_in_Dielectric}
\bibinfo{author}{\bibfnamefont{A.}~\bibnamefont{Schiffrin}},
  \bibinfo{author}{\bibfnamefont{T.}~\bibnamefont{Paasch-Colberg}},
  \bibinfo{author}{\bibfnamefont{N.}~\bibnamefont{Karpowicz}},
  \bibinfo{author}{\bibfnamefont{V.}~\bibnamefont{Apalkov}},
  \bibinfo{author}{\bibfnamefont{D.}~\bibnamefont{Gerster}},
  \bibinfo{author}{\bibfnamefont{S.}~\bibnamefont{Muhlbrandt}},
  \bibinfo{author}{\bibfnamefont{M.}~\bibnamefont{Korbman}},
  \bibinfo{author}{\bibfnamefont{J.}~\bibnamefont{Reichert}},
  \bibinfo{author}{\bibfnamefont{M.}~\bibnamefont{Schultze}},
  \bibinfo{author}{\bibfnamefont{S.}~\bibnamefont{Holzner}},
  \bibnamefont{et~al.}, \bibinfo{journal}{Nature}
  \textbf{\bibinfo{volume}{493}}, \bibinfo{pages}{70} (\bibinfo{year}{2012}).

\bibitem[{\citenamefont{Kelardeh et~al.}(2015)\citenamefont{Kelardeh, Apalkov,
  and Stockman}}]{Stockman_et_al_PRB_2015_Graphene_in_Strong_Field}
\bibinfo{author}{\bibfnamefont{H.~K.} \bibnamefont{Kelardeh}},
  \bibinfo{author}{\bibfnamefont{V.}~\bibnamefont{Apalkov}}, \bibnamefont{and}
  \bibinfo{author}{\bibfnamefont{M.~I.} \bibnamefont{Stockman}},
  \bibinfo{journal}{Phys. Rev. B} \textbf{\bibinfo{volume}{91}},
  \bibinfo{pages}{045439} (\bibinfo{year}{2015}).

\bibitem[{\citenamefont{Schultze et~al.}(2012)\citenamefont{Schultze,
  Bothschafter, Sommer, Holzner, Schweinberger, Fiess, Hofstetter, Kienberger,
  Apalkov, Yakovlev
  et~al.}}]{Schultze_et_al_Nature_2012_Controlling_Dielectrics}
\bibinfo{author}{\bibfnamefont{M.}~\bibnamefont{Schultze}},
  \bibinfo{author}{\bibfnamefont{E.~M.} \bibnamefont{Bothschafter}},
  \bibinfo{author}{\bibfnamefont{A.}~\bibnamefont{Sommer}},
  \bibinfo{author}{\bibfnamefont{S.}~\bibnamefont{Holzner}},
  \bibinfo{author}{\bibfnamefont{W.}~\bibnamefont{Schweinberger}},
  \bibinfo{author}{\bibfnamefont{M.}~\bibnamefont{Fiess}},
  \bibinfo{author}{\bibfnamefont{M.}~\bibnamefont{Hofstetter}},
  \bibinfo{author}{\bibfnamefont{R.}~\bibnamefont{Kienberger}},
  \bibinfo{author}{\bibfnamefont{V.}~\bibnamefont{Apalkov}},
  \bibinfo{author}{\bibfnamefont{V.~S.} \bibnamefont{Yakovlev}},
  \bibnamefont{et~al.}, \bibinfo{journal}{Nature}
  \textbf{\bibinfo{volume}{493}}, \bibinfo{pages}{75} (\bibinfo{year}{2012}).

\bibitem[{\citenamefont{Liu et~al.}(2011{\natexlab{d}})\citenamefont{Liu,
  Jiang, and
  Yao}}]{Liu_Jiang_Yao_PhysRevB.84_2011_Spin_Orbit_in_Germanium_and_Tin}
\bibinfo{author}{\bibfnamefont{C.-C.} \bibnamefont{Liu}},
  \bibinfo{author}{\bibfnamefont{H.}~\bibnamefont{Jiang}}, \bibnamefont{and}
  \bibinfo{author}{\bibfnamefont{Y.}~\bibnamefont{Yao}},
  \bibinfo{journal}{Phys. Rev. B} \textbf{\bibinfo{volume}{84}},
  \bibinfo{pages}{195430} (\bibinfo{year}{2011}{\natexlab{d}}).

\bibitem[{\citenamefont{Liu et~al.}(2011{\natexlab{e}})\citenamefont{Liu, Feng,
  and
  Yao}}]{Liu_Feng_Yao_PhysRevLett.107_2011_Quantum_Spin_Hall_Effect_in_Solicen%
e}
\bibinfo{author}{\bibfnamefont{C.-C.} \bibnamefont{Liu}},
  \bibinfo{author}{\bibfnamefont{W.}~\bibnamefont{Feng}}, \bibnamefont{and}
  \bibinfo{author}{\bibfnamefont{Y.}~\bibnamefont{Yao}},
  \bibinfo{journal}{Phys. Rev. Lett.} \textbf{\bibinfo{volume}{107}},
  \bibinfo{pages}{076802} (\bibinfo{year}{2011}{\natexlab{e}}).

\bibitem[{\citenamefont{Apalkov and
  Chakraborty}(2014)}]{Tunability_FQHE_dirac_materials_PRB_2014}
\bibinfo{author}{\bibfnamefont{V.~M.} \bibnamefont{Apalkov}} \bibnamefont{and}
  \bibinfo{author}{\bibfnamefont{T.}~\bibnamefont{Chakraborty}},
  \bibinfo{journal}{Phys. Rev. B} \textbf{\bibinfo{volume}{90}},
  \bibinfo{pages}{245108} (\bibinfo{year}{2014}).

\bibitem[{\citenamefont{Wallace}(1947)}]{Graphene_Wallace_PR_1947}
\bibinfo{author}{\bibfnamefont{P.~R.} \bibnamefont{Wallace}},
  \bibinfo{journal}{Phys. Rev.} \textbf{\bibinfo{volume}{71}},
  \bibinfo{pages}{622} (\bibinfo{year}{1947}).

\bibitem[{\citenamefont{Slonczewski and Weiss}(1958)}]{Graphene_Weiss_PR_1958}
\bibinfo{author}{\bibfnamefont{J.~C.} \bibnamefont{Slonczewski}}
  \bibnamefont{and} \bibinfo{author}{\bibfnamefont{P.~R.} \bibnamefont{Weiss}},
  \bibinfo{journal}{Phys. Rev.} \textbf{\bibinfo{volume}{109}},
  \bibinfo{pages}{272} (\bibinfo{year}{1958}).

\bibitem[{\citenamefont{Saito et~al.}(1998)\citenamefont{Saito, Dresselhaus,
  and Dresselhaus}}]{graphene_Dresselhaus_1998}
\bibinfo{author}{\bibfnamefont{R.}~\bibnamefont{Saito}},
  \bibinfo{author}{\bibfnamefont{G.}~\bibnamefont{Dresselhaus}},
  \bibnamefont{and}
  \bibinfo{author}{\bibfnamefont{M.}~\bibnamefont{Dresselhaus}},
  \emph{\bibinfo{title}{Physical Properties of Carbon Nanotubes}}
  (\bibinfo{publisher}{Imperial College Press}, \bibinfo{address}{London},
  \bibinfo{year}{1998}).

\bibitem[{\citenamefont{Reich et~al.}(2004)\citenamefont{Reich, Thomsen, and
  Maultzsch}}]{Carbon_nanotubes_2004}
\bibinfo{author}{\bibfnamefont{S.}~\bibnamefont{Reich}},
  \bibinfo{author}{\bibfnamefont{C.}~\bibnamefont{Thomsen}}, \bibnamefont{and}
  \bibinfo{author}{\bibfnamefont{J.}~\bibnamefont{Maultzsch}},
  \emph{\bibinfo{title}{Carbon Nanotubes}} (\bibinfo{publisher}{Wiley-VCH},
  \bibinfo{address}{Weinheim}, \bibinfo{year}{2004}).

\bibitem[{\citenamefont{Hwang and
  Sarma}(2008)}]{Hwang_Das_Sarma_PRB_2008_Graphene_Relaxation_Time}
\bibinfo{author}{\bibfnamefont{E.~H.} \bibnamefont{Hwang}} \bibnamefont{and}
  \bibinfo{author}{\bibfnamefont{S.} \bibnamefont{DasSarma}},
  \bibinfo{journal}{Phys. Rev. B} \textbf{\bibinfo{volume}{77}},
  \bibinfo{pages}{195412} (\bibinfo{year}{2008}).

\bibitem[{\citenamefont{Breusing et~al.}(2011)\citenamefont{Breusing, Kuehn,
  Winzer, Malic, Milde, Severin, Rabe, Ropers, Knorr, and
  Elsaesser}}]{ultrafast_dynamics_graphene_PRB_2011}
\bibinfo{author}{\bibfnamefont{M.}~\bibnamefont{Breusing}},
  \bibinfo{author}{\bibfnamefont{S.}~\bibnamefont{Kuehn}},
  \bibinfo{author}{\bibfnamefont{T.}~\bibnamefont{Winzer}},
  \bibinfo{author}{\bibfnamefont{E.}~\bibnamefont{Malic}},
  \bibinfo{author}{\bibfnamefont{F.}~\bibnamefont{Milde}},
  \bibinfo{author}{\bibfnamefont{N.}~\bibnamefont{Severin}},
  \bibinfo{author}{\bibfnamefont{J.~P.} \bibnamefont{Rabe}},
  \bibinfo{author}{\bibfnamefont{C.}~\bibnamefont{Ropers}},
  \bibinfo{author}{\bibfnamefont{A.}~\bibnamefont{Knorr}}, \bibnamefont{and}
  \bibinfo{author}{\bibfnamefont{T.}~\bibnamefont{Elsaesser}},
  \bibinfo{journal}{Phys. Rev. B} \textbf{\bibinfo{volume}{83}},
  \bibinfo{pages}{153410} (\bibinfo{year}{2011}).

\bibitem[{\citenamefont{Malic et~al.}(2011)\citenamefont{Malic, Winzer, Bobkin,
  and Knorr}}]{theory_absorption_ultrafast_kinetics_graphene_PRB_2011}
\bibinfo{author}{\bibfnamefont{E.}~\bibnamefont{Malic}},
  \bibinfo{author}{\bibfnamefont{T.}~\bibnamefont{Winzer}},
  \bibinfo{author}{\bibfnamefont{E.}~\bibnamefont{Bobkin}}, \bibnamefont{and}
  \bibinfo{author}{\bibfnamefont{A.}~\bibnamefont{Knorr}},
  \bibinfo{journal}{Phys. Rev. B} \textbf{\bibinfo{volume}{84}},
  \bibinfo{pages}{205406} (\bibinfo{year}{2011}).

\bibitem[{\citenamefont{Brida et~al.}(2013)\citenamefont{Brida, Tomadin,
  Manzoni, Kim, Lombardo, Milana, Nair, Novoselov, Ferrari, Cerullo
  et~al.}}]{Ultrafast_collinear_scattering_graphene_nat_comm_2013}
\bibinfo{author}{\bibfnamefont{D.}~\bibnamefont{Brida}},
  \bibinfo{author}{\bibfnamefont{A.}~\bibnamefont{Tomadin}},
  \bibinfo{author}{\bibfnamefont{C.}~\bibnamefont{Manzoni}},
  \bibinfo{author}{\bibfnamefont{Y.~J.} \bibnamefont{Kim}},
  \bibinfo{author}{\bibfnamefont{A.}~\bibnamefont{Lombardo}},
  \bibinfo{author}{\bibfnamefont{S.}~\bibnamefont{Milana}},
  \bibinfo{author}{\bibfnamefont{R.~R.} \bibnamefont{Nair}},
  \bibinfo{author}{\bibfnamefont{K.~S.} \bibnamefont{Novoselov}},
  \bibinfo{author}{\bibfnamefont{A.~C.} \bibnamefont{Ferrari}},
  \bibinfo{author}{\bibfnamefont{G.}~\bibnamefont{Cerullo}},
  \bibnamefont{et~al.}, \bibinfo{journal}{Nat. Commun.}
  \textbf{\bibinfo{volume}{4}} (\bibinfo{year}{2013}).

\bibitem[{\citenamefont{Gierz et~al.}(2013)\citenamefont{Gierz, Petersen,
  Mitrano, Cacho, Turcu, Springate, StÃ¶hr, KÃ¶hler, Starke, and
  Cavalleri}}]{non-equilibrium_Dirac_carrier_distributions_graphene_nat_materi%
als_2013}
\bibinfo{author}{\bibfnamefont{I.}~\bibnamefont{Gierz}},
  \bibinfo{author}{\bibfnamefont{J.~C.} \bibnamefont{Petersen}},
  \bibinfo{author}{\bibfnamefont{M.}~\bibnamefont{Mitrano}},
  \bibinfo{author}{\bibfnamefont{C.}~\bibnamefont{Cacho}},
  \bibinfo{author}{\bibfnamefont{I.~C.~E.} \bibnamefont{Turcu}},
  \bibinfo{author}{\bibfnamefont{E.}~\bibnamefont{Springate}},
  \bibinfo{author}{\bibfnamefont{A.}~\bibnamefont{StÃ¶hr}},
  \bibinfo{author}{\bibfnamefont{A.}~\bibnamefont{KÃ¶hler}},
  \bibinfo{author}{\bibfnamefont{U.}~\bibnamefont{Starke}}, \bibnamefont{and}
  \bibinfo{author}{\bibfnamefont{A.}~\bibnamefont{Cavalleri}},
  \bibinfo{journal}{Nat. Mater.} \textbf{\bibinfo{volume}{12}},
  \bibinfo{pages}{1119} (\bibinfo{year}{2013}), ISSN \bibinfo{issn}{1476-1122}.

\bibitem[{\citenamefont{Tomadin et~al.}(2013)\citenamefont{Tomadin, Brida,
  Cerullo, Ferrari, and
  Polini}}]{Nonequilibrium_dynamics_photoexcited_electrons_graphene_PRB_2013}
\bibinfo{author}{\bibfnamefont{A.}~\bibnamefont{Tomadin}},
  \bibinfo{author}{\bibfnamefont{D.}~\bibnamefont{Brida}},
  \bibinfo{author}{\bibfnamefont{G.}~\bibnamefont{Cerullo}},
  \bibinfo{author}{\bibfnamefont{A.~C.} \bibnamefont{Ferrari}},
  \bibnamefont{and} \bibinfo{author}{\bibfnamefont{M.}~\bibnamefont{Polini}},
  \bibinfo{journal}{Phys. Rev. B} \textbf{\bibinfo{volume}{88}},
  \bibinfo{pages}{035430} (\bibinfo{year}{2013}).

\bibitem[{\citenamefont{Krausz and
  Stockman}(2014)}]{Krausz_Stockman_Nature_Photonics_2014_Attosecond_Review}
\bibinfo{author}{\bibfnamefont{F.}~\bibnamefont{Krausz}} \bibnamefont{and}
  \bibinfo{author}{\bibfnamefont{M.~I.} \bibnamefont{Stockman}},
  \bibinfo{journal}{Nat. Phot.} \textbf{\bibinfo{volume}{8}},
  \bibinfo{pages}{205} (\bibinfo{year}{2014}).

\bibitem[{\citenamefont{Schultze et~al.}(2014)\citenamefont{Schultze,
  Ramasesha, Pemmaraju, Sato, Whitmore, Gandman, Prell, Borja, Prendergast,
  Yabana et~al.}}]{Schultze-2014-Attosecond_band-gap}
\bibinfo{author}{\bibfnamefont{M.}~\bibnamefont{Schultze}},
  \bibinfo{author}{\bibfnamefont{K.}~\bibnamefont{Ramasesha}},
  \bibinfo{author}{\bibfnamefont{C.~D.} \bibnamefont{Pemmaraju}},
  \bibinfo{author}{\bibfnamefont{S.~A.} \bibnamefont{Sato}},
  \bibinfo{author}{\bibfnamefont{D.}~\bibnamefont{Whitmore}},
  \bibinfo{author}{\bibfnamefont{A.}~\bibnamefont{Gandman}},
  \bibinfo{author}{\bibfnamefont{J.~S.} \bibnamefont{Prell}},
  \bibinfo{author}{\bibfnamefont{L.~J.} \bibnamefont{Borja}},
  \bibinfo{author}{\bibfnamefont{D.}~\bibnamefont{Prendergast}},
  \bibinfo{author}{\bibfnamefont{K.}~\bibnamefont{Yabana}},
  \bibnamefont{et~al.}, \bibinfo{journal}{Science}
  \textbf{\bibinfo{volume}{346}}, \bibinfo{pages}{1348} (\bibinfo{year}{2014}).

\bibitem[{\citenamefont{Wannier}(1960)}]{Wannier_PR_1960_Wannier_States_in_Str%
ong_Fields}
\bibinfo{author}{\bibfnamefont{G.~H.} \bibnamefont{Wannier}},
  \bibinfo{journal}{Phys. Rev.} \textbf{\bibinfo{volume}{117}},
  \bibinfo{pages}{432} (\bibinfo{year}{1960}).

\bibitem[{\citenamefont{Houston}(1940)}]{Houston_PR_1940_Electron_Acceleration%
_in_Lattice}
\bibinfo{author}{\bibfnamefont{W.~V.} \bibnamefont{Houston}},
  \bibinfo{journal}{Phys. Rev.} \textbf{\bibinfo{volume}{57}},
  \bibinfo{pages}{184} (\bibinfo{year}{1940}).

\bibitem[{\citenamefont{Landau and
  Lifshitz}(1965)}]{Landau_Lifshitz_Quantum_Mechanics:1965}
\bibinfo{author}{\bibfnamefont{L.~D.} \bibnamefont{Landau}} \bibnamefont{and}
  \bibinfo{author}{\bibfnamefont{E.~M.} \bibnamefont{Lifshitz}},
  \emph{\bibinfo{title}{Quantum Mechanics: Non-Relativistic Theory}}
  (\bibinfo{publisher}{Pergamon Press}, \bibinfo{address}{Oxford and New York},
  \bibinfo{year}{1965}).

\bibitem[{\citenamefont{Apalkov and
  Stockman}(2012)}]{Apalkov_Stockman_PRB_2012_Strong_Field_Reflection}
\bibinfo{author}{\bibfnamefont{V.}~\bibnamefont{Apalkov}} \bibnamefont{and}
  \bibinfo{author}{\bibfnamefont{M.~I.} \bibnamefont{Stockman}},
  \bibinfo{journal}{Phys. Rev. B} \textbf{\bibinfo{volume}{86}},
  \bibinfo{pages}{165118} (\bibinfo{year}{2012}).

\bibitem[{\citenamefont{Schiffrin et~al.}(2014)\citenamefont{Schiffrin,
  Paasch-Colberg, Karpowicz, Apalkov, Gerster, Muhlbrandt, Korbman, Reichert,
  Schultze, Holzner
  et~al.}}]{Schiffrin_at_al_Nature_2014_Current_in_Dielectric_Addendum}
\bibinfo{author}{\bibfnamefont{A.}~\bibnamefont{Schiffrin}},
  \bibinfo{author}{\bibfnamefont{T.}~\bibnamefont{Paasch-Colberg}},
  \bibinfo{author}{\bibfnamefont{N.}~\bibnamefont{Karpowicz}},
  \bibinfo{author}{\bibfnamefont{V.}~\bibnamefont{Apalkov}},
  \bibinfo{author}{\bibfnamefont{D.}~\bibnamefont{Gerster}},
  \bibinfo{author}{\bibfnamefont{S.}~\bibnamefont{Muhlbrandt}},
  \bibinfo{author}{\bibfnamefont{M.}~\bibnamefont{Korbman}},
  \bibinfo{author}{\bibfnamefont{J.}~\bibnamefont{Reichert}},
  \bibinfo{author}{\bibfnamefont{M.}~\bibnamefont{Schultze}},
  \bibinfo{author}{\bibfnamefont{S.}~\bibnamefont{Holzner}},
  \bibnamefont{et~al.}, \bibinfo{journal}{Nature}
  \textbf{\bibinfo{volume}{507}}, \bibinfo{pages}{386} (\bibinfo{year}{2014}).

\bibitem[{\citenamefont{Landau}(1946)}]{Landau_J_Phys_1946_Landau_Damping}
\bibinfo{author}{\bibfnamefont{L.~D.} \bibnamefont{Landau}},
  \bibinfo{journal}{Journal of Physics} \textbf{\bibinfo{volume}{10}},
  \bibinfo{pages}{25} (\bibinfo{year}{1946}).

\bibitem[{\citenamefont{Paasch-Colberg
  et~al.}(2014)\citenamefont{Paasch-Colberg, Schiffrin, Karpowicz, Kruchinin,
  Ozge, Keiber, Razskazovskaya, Muhlbrandt, Alnaser, Kubel
  et~al.}}]{Stockman_et_al_Nat_Phot_2013_CEP_Detector}
\bibinfo{author}{\bibfnamefont{T.}~\bibnamefont{Paasch-Colberg}},
  \bibinfo{author}{\bibfnamefont{A.}~\bibnamefont{Schiffrin}},
  \bibinfo{author}{\bibfnamefont{N.}~\bibnamefont{Karpowicz}},
  \bibinfo{author}{\bibfnamefont{S.}~\bibnamefont{Kruchinin}},
  \bibinfo{author}{\bibfnamefont{S.}~\bibnamefont{Ozge}},
  \bibinfo{author}{\bibfnamefont{S.}~\bibnamefont{Keiber}},
  \bibinfo{author}{\bibfnamefont{O.}~\bibnamefont{Razskazovskaya}},
  \bibinfo{author}{\bibfnamefont{S.}~\bibnamefont{Muhlbrandt}},
  \bibinfo{author}{\bibfnamefont{A.}~\bibnamefont{Alnaser}},
  \bibinfo{author}{\bibfnamefont{M.}~\bibnamefont{Kubel}},
  \bibnamefont{et~al.}, \bibinfo{journal}{Nat. Phot.}
  \textbf{\bibinfo{volume}{8}}, \bibinfo{pages}{214} (\bibinfo{year}{2014}).

\end{thebibliography}

\end{document}